# Numerical Investigation of Supercritical Combustion Dynamics in a Multi-Element LOx-Methane Combustor Using Flamelet-Generated Manifold Approach

Abhishek Sharma[1,2], Ashoke De[2,3*], S.Sunil Kumar[1]

[1]Liquid Propulsion Systems Center, ISRO, Valiamala, 695547, Thiruvananthapuram, India
[2]Department of Aerospace Engineering, Indian Institute of Technology Kanpur, 208016, Kanpur, India
[3]Department of Sustainable Energy Engineering, Indian Institute of Technology Kanpur, 208016, Kanpur, India

The article investigates liquid oxygen (LOx)-methane supercritical combustion dynamics in a multi-element rocket scale combustor using large eddy simulation (LES). A complex framework of real gas thermodynamics and flamelet generated manifold (FGM) combustion model is invoked to simulate transcritical oxygen injection and supercritical methane combustion. A benchmark Mascotte chamber, Rocket Combustion Modelling (RCM) Test Case, i.e. RCM-3 (V04)/G2 test case, is used to validate the real gas FGM model in the LES framework. The validation study accurately reproduces experimental flame structure and OH concentration, demonstrating the FGM model's importance in incorporating finite rate kinetics in LOx-methane combustion. Subsequently, the numerical framework investigates a specially designed multi-element combustion chamber featuring seven bidirectional swirl coaxial injectors. The analyses capture the complex hydrodynamics and combustion dynamics associated with multiple swirl injectors operating at supercritical pressure, effectively demonstrating the initiation of transverse acoustic waves and examining the effect of local sound speed on the evolution of acoustic modes in the combustor. The dominant frequency modes shed light on understanding the role of injectors in enhanced combustor dynamics. Spectral analysis reveals the interplay of the upstream injector and chamber acoustics due to possible frequency coupling. The results also highlight the effect of fuel injection temperature on the stability of the combustor, revealing a violent dynamic activity for lower fuel injection temperature associated with the longitudinal acoustic mode of the combustor. The investigation appropriately reproduces self-sustained limit cycle oscillations at lower fuel injection temperatures and corroborates the conventional understanding of combustor instability.

## I. INTRODUCTION

The complex nature of transcritical injection and supercritical LOx-methane combustion has been widely studied in recent years. The complex thermodynamic and transport phenomenon associated with injection and combustion in such high-pressure systems is well reported in reference[1]. It reviews the process and challenges associated with experimental and numerical modeling of high-pressure LOx-methane rocket engines. The high-pressure operation makes the engine susceptible to oscillations, known as combustion instability/dynamics, which can result in significant heat transfer and potential damage to the engine. Many

---

Author to whom correspondence should be addressed.  Electronic mail:  ashoke@iitk.ac.in

researchers have contributed to understanding combustion dynamics in liquid rocket engines: Crocco and Cheng[2,3], Zinn[4], Oefelein and Yang[5], Yang and Anderson[6], Yang[7], Culick[8], Sirignano[9,10], Candel[11], Poinsot[12]. The mechanism of combustion dynamics in liquid rocket engines is complicated due to the nonlinear phenomenon associated with the interactions of multiple injectors, injector-chamber, and chamber-upstream feed systems. The understanding of combustion dynamics under supercritical conditions is still unclear and presents a significant challenge, both in terms of experimental investigations and computational analysis. Our earlier work[1] presented a detailed review of numerical investigation on stable characteristics of supercritical LOx-methane combustion. In this section, we provide a brief review of the unsteady flame dynamic aspects of supercritical LOx-methane flames. Although many numerical studies[13–16] have been conducted to study self-excited instability in gaseous oxygen (GOx)-methane single-element combustor, only a few are dedicated to explore supercritical LOx-methane combustion. Zong et al.[17,18] utilized LES to investigate LOx-methane supercritical combustion in a single shear coaxial injector with a laminar flamelet model. Their study identified strong vortex shedding due to significant density differences between LOx and gaseous methane. The driving force behind the flow dynamics was observed to be the local thermodynamic state and its impact on the thermophysical properties. Huo et al.[19] conducted a detailed investigation of LOx-methane supercritical combustion at a single injector element to determine the effect of injector geometry on flame dynamics. Schmitt[20] studied LOx-methane flame dynamics in a single-element Mascotte chamber with a steady laminar flamelet model in conjunction with the Soave Redlich-Kwong (SRK) equation of state (EOS) in the LES framework. Muller[21], Traxinger[22], and Zips et al.[23] developed a real gas efficient flamelet-based LES framework for the Mascotte G2 test case. Xiong et al.[24,25] recently studied the methane-oxygen combustion instability in a shear coaxial injector combustor, performing simulations at a chamber pressure of 150-200 bar with an ideal gas equation of state in the detached eddy simulation framework. A single-step reaction finite-rate chemistry model is used to simulate methane-oxygen combustion. The simulation captured spontaneous and triggered longitudinal and tangential mode instability. However, there is an impact on the accuracy of results due to using the ideal gas equation and ignoring the turbulent fluctuations at such high-pressure conditions. Wang et al.[26,27] further elaborated dynamics associated with swirl coaxial injectors. They presented a comprehensive investigation of swirl injector flow dynamics at supercritical conditions. LES framework with a steady flamelet combustion model is used to study the injection and combustion dynamics in a single-element injector domain. The study reported many numerical tests conducted at single element level to understand the role of swirl, LOx post thickness, recess, fuel end taper, etc. The flow field of the swirl injector and underlying instability mechanisms such as shear-layer, helical, centrifugal, and acoustic instabilities are thoroughly investigated, and hydrodynamic instability was shown to be the primary cause of flow oscillations in the injector. Recent literature[28–32] highlights many numerical investigations on supercritical combustion dynamics conducted under simplified geometric conditions, with most of the LES studies only conducted at a single-element level. Even though complex non-ideal thermodynamic and transport models are invoked in high-fidelity numerical frameworks, the multi-element flame dynamics effect is not accounted for, and the multi-element flame-flame and flame-wall interactions are often ignored. Urbano et al.[12] investigated a multi-element real gas simulation of the German Aerospace Center (DLR) combustor[33], operating on LOx-hydrogen propellants. They considered a shear coaxial injector element, while a steady flamelet approach was invoked to model hydrogen-oxygen chemistry. The simulation captured limit cycle oscillations when triggered with high amplitude pressure disturbance.

Existing literature suggests very few studies have been conducted to comprehensively understand the dynamics of LOx-methane supercritical combustion, especially on a multi-element scale. The dynamics of multiple swirl coaxial injectors, injector-chamber interactions, and the effect of injection conditions on the stability of the combustor are yet to be explored. Although a detailed assessment of combustion models in our earlier work[1] demonstrated the usefulness of Flamelet generated manifold model (FGM)[34] for LOx-methane supercritical flames, implementing the FGM model in a high-fidelity numerical framework for combustion dynamics is required. The unsteady flame dynamics can influence the stability of the combustor, and the role of combustion closure, such as FGM, which can accommodate transient effects in the diffusion flame framework, must be explored. The main goal of this study is to implement a validated numerical methodology to simulate supercritical LOx-methane combustion dynamics in a multi-element rocket scale combustor. The focus is to incorporate complex real gas thermodynamics and FGM model into a scale-resolved LES framework that can be suitably applied on a designed 7-element combustor at nominal operating conditions. To our knowledge, no study is reported in the literature that simulates LOx-methane supercritical combustion dynamics in a multi-element combustor using the FGM approach. The study has been further extended to study the impact of the off-nominal condition, i.e., lower fuel injection temperature, on the stability of the 7-element combustor.

Conventionally, lower fuel injection temperature triggers combustion instability[3], and there can be a threshold limit for fuel temperature below which the combustor can become unstable. It is necessary to ascertain this limit through experimentation or computations for any new combustor design. Computational studies to investigate the impact of lower fuel injection temperatures on stability are rare and have primarily focused on laboratory-scale combustors. Literature reports few attempts in this direction. Schmitt et al.[36] simulated DLR combustor[37] tests at 90K and 45K hydrogen inlet temperatures, respectively. The study showed dynamic activity in the combustor compared to stable operation observed in the tests at an injection temperature of 45K. This difference in behavior was attributed to the dynamic fluid conditions that prevailed during the test, as opposed to the static inputs used in simulations. Schmitt et al.[38] also studied JAXA shear coaxial single-element long combustor using an infinitely fast chemistry model at varied hydrogen injection temperatures. Dominant chamber frequency was captured in case of lower hydrogen injection temperature, even though the amplitude was lower than the test data. A recent two-dimensional computational study by Ota et al.[39] investigated the effect of lower hydrogen temperature on the stability of a single-element $H_2$-$O_2$ combustor. It shows stable combustion for fuel injection at 300K and severe pressure oscillations at a low fuel injection temperature of 80K. It shows the onset of the acoustic wave corresponding to the first longitudinal mode of the combustor. Harvazinski et al.[40] studied the impact of lower oxidizer temperature on the stability of a continuously variable resonance combustor, CVRC[41]. It shows the lower oxidizer temperature increases the amplitude of the pressure oscillations, which was attributed to an increase in the combustion recovery time after the fuel cut-off event. Until now, the existing literature lacks numerical investigations into the impact of reduced fuel injection temperatures on the stability of high-pressure LOx-methane flames, particularly in a multi-element rocket scale combustor. The goal is to conduct computations at a reduced methane injection temperature and to compare combustion dynamics for two different fuel inlet temperature conditions.

The primary motivation of the present work is to address a significant gap in the field by examining the combustion dynamics of LOx-methane flames in a rocket-scale combustor featuring multiple injectors. Our review of existing literature has revealed a noticeable lack of comprehensive studies concerning the dynamics of LOx-methane supercritical combustion, particularly at a multi-element scale. Until now, most investigations have been limited to the single injector element level, ignoring flame-flame and flame-wall interactions. The injectors considered in literature studies are plane shear coaxial injectors, with no study conducted with multiple swirl coaxial injectors. Our study explores the multiple aspects of LOx-methane combustion and presents the dynamics of multiple swirl coaxial injectors, injector-chamber interactions, and the effect of injection conditions on the stability of the combustor, thereby bridging this research gap.

The paper comprises three main sections outlined as Section II provides a brief overview of the theoretical details and FGM formulation for LOx-methane flames. Section III presents a validation study on the well-known Mascotte G2 test case of Singla et al.[42]. The validation study assesses the appropriateness of the numerical framework. Section IV of the article presents multi-element combustor dynamics and illustrates the key physics of propellant injection, mixing, vorticity, and flame characteristics at supercritical operating conditions. The influence of injector dynamics on the combustor stability is investigated and presented through detailed spectral analysis. This section also highlights the impact of fuel injection temperature on the stability of the combustor, where the case of lower fuel injection temperature is simulated, and a detailed comparative analysis is conducted with a stable operating case. The numerical framework captures the self-sustained limit cycle oscillations in LOx-methane supercritical flames and sheds light on the conventional understanding of instability occurrence at lower fuel injection temperatures.

## II. THEORETICAL DETAILS AND FGM FORMULATION

The work employs a LES framework incorporating real gas thermodynamics and a high-fidelity combustion model to resolve turbulence and combustion scales. The best-known practices evolved through earlier work[16] on the simulation of self-excited instability in continuously variable resonance combustors, CVRC, are employed in this study. The framework solves compressible Favre-filtered continuity, momentum, energy, and other governing equations. Appendix A provides the details of the governing equations used herein. Our earlier work[1] presented a detailed evaluation of thermodynamic and transport properties at supercritical operating conditions. It showcased the Soave-Redlich-Kwong (SRK)[43] real gas model to accurately compute thermodynamic states for both methane and oxygen over an entire range of temperature and pressures considered in the present study. The transport properties are calculated as per the proposed method of Chung et al.[44]. A brief overview of mixture thermodynamic models is provided in Appendix A, although details can be referred from our earlier work[1]. A brief description of FGM formulation for LOx-methane flames is provided in the following subsection.

### A. Flamelet Generated Manifold (FGM) Formulation

FGM is a dimension-reduced approach that combines steady flamelet and intrinsic low-dimensional manifold (ILDM) formulations to overcome the limitations associated with each approach. This study utilizes the formulation proposed by Oijen and Goey et al.[34], which assumes that the thermo-chemical states in a turbulent flame are similar to those in a laminar flame, and control variables like mixture fraction and reaction progress can parameterize the flame. The significant advantage of this approach is that chemical kinetic effects can be incorporated into turbulent flames by solving the transport equation of the reaction progress variable. Our earlier study[1] in the Reynolds-Averaged Navier-Stokes (RANS) framework reveals the fundamental appropriateness of the FGM model to capture LOx methane flames at supercritical conditions. However, a brief overview of FGM formulation is provided in Appendix A. As described in our previous work[1], the unsteady flamelet equations are solved to generate diffusion flamelets. FGM creates a two-dimensional manifold in mixture fraction-progress variable space, where equation (A18) maps the flamelet species field calculated at different scalar dissipations into a normalized progress variable ($c$). This study discretizes mixture fraction and progress variable space with 100 grid points to capture the sharp gradients in flame temperature even away from $c = 1$ condition. Our previous study[1] showcased the utility of the 6 steps (small) JL[45] mechanism for correctly predicting OH mass fraction in the FGM framework. However, in the current work, we have invoked the detailed mechanism by Kong et al.[46] to generate multiple flamelets. This detailed mechanism accounts for the impact of multiple species at high-pressure operating conditions. Figures 1 and 2 depict FGM flamelets generated at 60 bar conditions. Figure 1(a) displays flamelet temperature varying with mixture fraction and progress variable. The maximum temperature occurs in burnt flamelet at $c=1$ and a mixture fraction value of 0.2132. The flamelet temperature from $c=1$ to lower values of $c$ describes the flamelet manifold considered in this study. Figure 1(b) presents the species mass fraction in the burnt flamelet. A peak in $H_2O$, $CO_2$, and OH mass fraction is close to the stoichiometric mixture fraction, whereas CO peaks in the fuel-rich flame region.

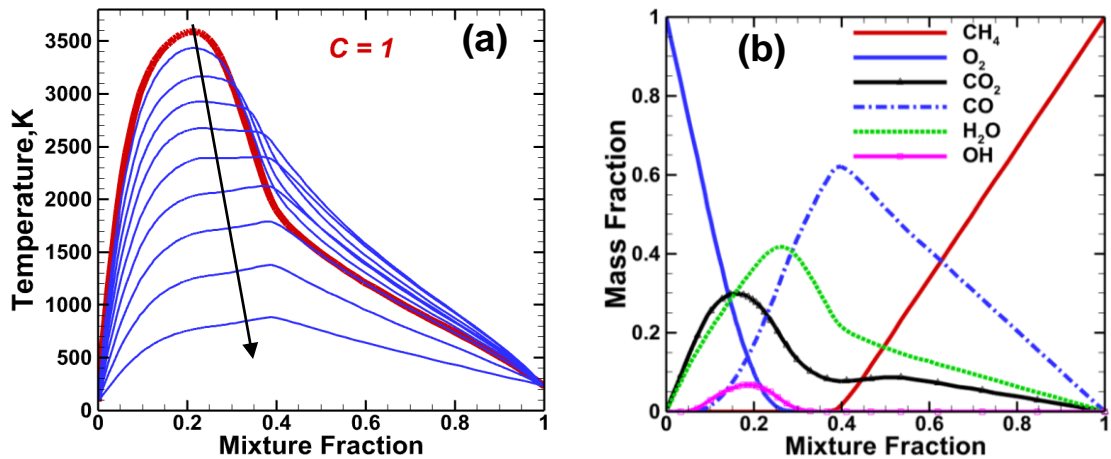

**FIG.1.** (a) Temperature (K), (b) Mass fraction variation with mixture fraction ($f$)

Figure 2(a) shows temperature variation with progress variable at a particular value of mixture fraction. It reports the maximum temperature at $c=1$ for $f=0.1$ and 0.2, while peak temperature for $f = 0.4$ and 0.6 occurs at $c$ close to 0.5. To capture this peak temperature at higher values of $f$, a higher number of grid points are inserted even away from burnt and stoichiometric state, i.e. $c=1$ and $f=0.2132$, respectively. Figure 2(b) further illustrates the difference in peak $CO_2$ concentration close to and far from the stoichiometric mixture fraction (0.2132).

Multiple flamelets are generated appropriately to capture these variations with a higher number of $c$ and $f$ space grid points.

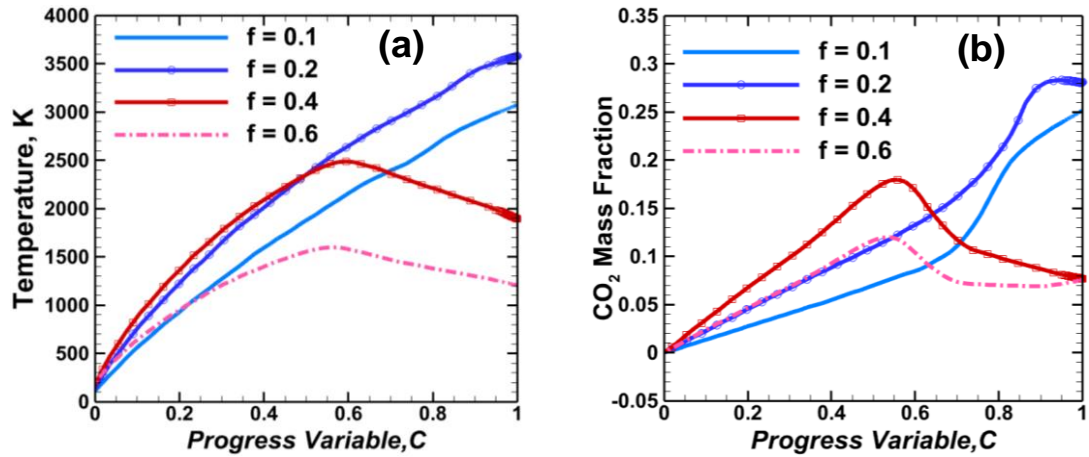

**FIG.2.** (a) Temperature (K), (b) $CO_2$ Mass fraction variation with progress variable ($c$)

In this study, turbulence-chemistry interaction is described using well known assumed shape probability density function (PDF) method. The average value of fluctuating scalars is computed by convoluting the flamelet profiles with the assumed-shape PDF. A brief description of the turbulence-chemistry interaction is provided in Appendix A. The mean quantities are tabulated in a look-up table against a specified value of 5 parameters, such as $\bar{f}$, $\overline{f'^2}$, mean $\bar{c}$, variance $\overline{c'^2}$ of progress variable respectively and mean enthalpy, $\bar{H}$. In this study, to reduce the computational cost of generating a five-dimensional table, it is assumed that the heat loss or gain by the system only has a negligible effect on species mass fraction, and mean scalars other than species mass fraction can be computed using the average progress variable. A different set of look-up tables for species mass fraction $\bar{\varphi}(\bar{f}, \overline{f'^2}, \bar{c}, \overline{c'^2})$ and remaining scalars affected by enthalpy change $\bar{\varphi}(\bar{f}, \overline{f'^2}, \bar{c}, \bar{H})$ is generated respectively.

All governing equations are implicitly filtered by the finite-volume methodology of ANSYS Fluent[47], with spatial discretization performed by second-order bounded schemes and time integration using the bounded second-order implicit method. A sufficiently small-time step size based on the acoustic CFL (ACFL) number is used to resolve the acoustic waves, maintain stability, and reduce numerical diffusion. We have performed all the simulations on 20 CPU compute nodes of a high-performance computing facility established at the liquid propulsion systems center (LPSC). 560 direct water-cooled Intel Broadwell processor cores are utilized to execute multiple cases.

## III. SOLVER VALIDATION

We have carried out a validation test of the computational framework on a well-known Mascotte chamber experimented by Candel and Singla et al.[11]. The facility was initially developed for hydrogen-oxygen combustion but was further extended to perform remarkable experimental investigations on LOx-methane combustion. It is a relevant benchmark case to test the numerical framework at rocket-relevant operating conditions. LOx-methane experiments on the Mascotte chamber include various operating regimes such as subcritical, trans-critical, and doubly trans-critical chamber conditions for a wide range of operating pressures (0.1 to 7MPa). The details on the Mascotte test facility and experimentation performed can be referred from our earlier work[1]. In this study, we have compared the test results from the RCM-3 (V04)/G2 test case. In the G2 test, the Mascotte chamber operates at a pressure of 5.61 MPa, with methane injected in a supercritical state while the oxygen is in a transcritical state at a temperature of 85 K. It uses a single shear coaxial injector, with the coaxial entry to methane along with the oxygen entry from the center. The test conditions are oxygen to fuel mixture ratio of 0.31, with a methane mass flow rate of 0.0444 kg/s and an oxygen mass flow rate of 0.1431 kg/s. Table 1 displays the operating conditions for the G2 test case, highlighting the high-density liquid-like condition for oxygen and the gaseous inlet condition for methane. Singla et al.[42] assessed OH and CH concentrations in the supercritical flame. In this study, we have used Abel transformed OH concentration image from the G2 test to compare with the current simulation results.

TABLE I. Operating condition for G2 case

| LOx/CH$_4$ –Mascotte G2 Case | | | | | | |
|---|---|---|---|---|---|---|
| P(MPa) | T$_{O2}$ (K) | T$_{CH4}$ (K) | U$_{O2}$(m/s) | U$_{CH4}$(m/s) | ρ$_{O2}$(kg/m$^3$) | ρ$_{CH4}$(kg/m$^3$) |
| 5.61 | 85 | 288 | 2.48 | 63.20 | 1180.08 | 41.79 |

The Mascotte chamber has a rectangular cross-section of a $50 \times 50$ mm$^2$ square duct and a length of 400mm, which ends with a converging-diverging nozzle. To reduce the computational expense of LES, we have invoked a truncated domain of 200mm in this study. Figure 3 displays a rectangular computational domain with a shear coaxial injector and representative flame interface at the entry. The geometric details of the injector and chamber are similar, as reported by Kim et al.[48] Figure 3 shows a zoomed view of the injector with geometric features like LOx post thickness and post divergence appropriately modeled without simplification.

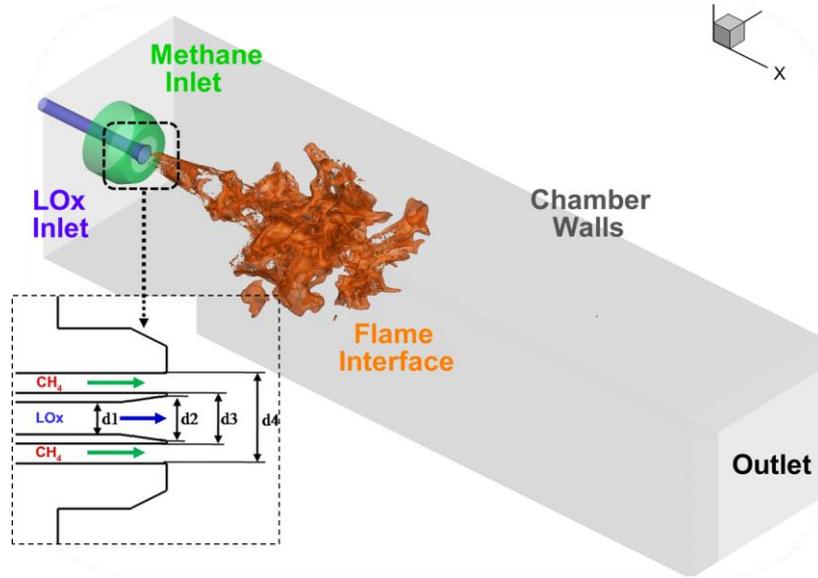

**FIG.3.** Computational domain of G2 test chamber (representative flame) with a zoomed view of the injector

Table 2 provides the geometric details of the G2 test case as mentioned in our previous RANS based work[1]. It displays injector dimensions with d1, d2, d3, and d4 parameters. The computational domain length is sufficient to contain the flame and avoid backflow effects from the outlet boundary. Methane and oxygen inlet are imposed with the mass flow inlet boundary conditions and outlet with pressure outlet boundary conditions. All walls are imposed with adiabatic and no-slip boundary conditions.

**TABLE II.** Geometric details of the G2 test case[48]

| Injector geometric details | Dimensions |
| --- | --- |
| LOx inner diameter (d1) | 3.6 mm |
| LOx outer diameter (d2) | 5.0 mm |
| LOx post divergence angle | $10^o$ |
| $CH_4$ inner diameter (d3) | 5.6 mm |
| $CH_4$ outer diameter (d4) | 10 mm |
| Chamber length | 200 mm |
| Chamber height | 50 mm |

The LES computational expense is reduced by providing the perturbed RANS data as an initial turbulent condition. The simulation is performed for over 50 flow-through times before initiating data sampling. We have tested the numerical framework on two grid levels to find the grid level that can resolve turbulent scales within a reasonable computation timeframe. This study uses two refined grids: fine (~3 million cells) and finer mesh (~ 6 million cells). A higher grid intensity in the flame region with a minimum cell size of 0.55mm and 0.25mm is used in fine and finer grids, respectively. The grid quality is checked by the LES resolution criterion defined as:

$$LES_{resolution} = \frac{k_{resolved}}{k_{resolved} + k_{SGS}} \qquad (1)$$

where $k_{resolved}$ is the resolved kinetic energy expressed as:

$$k_{resolved} = 0.5[(u - \bar{u})^2 + (v - \bar{v})^2 + (w - \bar{w})^2] \qquad (2)$$

with *u, v, w* the three components of the flow velocity, and the overbar denotes time-averaging.

$k_{SGS}$ expresses the unresolved sub-grid scale (SGS) turbulent kinetic energy, where $\mu_t$ is turbulent viscosity and $c_{var}$ is the dynamic constant.

$$k_{SGS} = 1.5\left(\frac{\mu_t}{\rho c_{var}\Delta_f}\right)^2, \text{constant } \Delta_f \text{ is calculated by } (Cell\ volume)^{1/3} \qquad (3)$$

The low value of modeled sub-grid scale viscosity in the core flow/flame region shows the suitability of the grid resolution, resolving a minimum of 80% of turbulent scales in the domain. Figure 4(a) displays the LES resolution of fine (top) and finer (bottom) mesh, with an overall LES resolution close to 1 in both grids. Figure 4(b) compares the axial mean temperature in both grids. It shows the temperature peak for both grids in a close match with an experimental band. The experimental band signifies the average flame location in the G2 test. Both the LES resolution map and temperature variation show that both grids can resolve turbulent scales with acceptable accuracy while reproducing primary flow and flame characteristics. We have chosen the fine grid for further analysis, considering the similarity of results in both grids and to optimize further the computational time required for sampling the LES data.

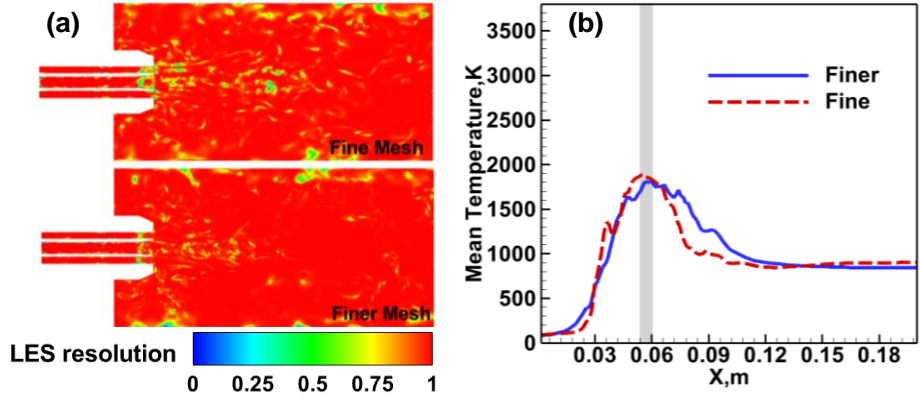

**FIG.4.** (a) LES resolution, (b) mean temperature plot for two grids

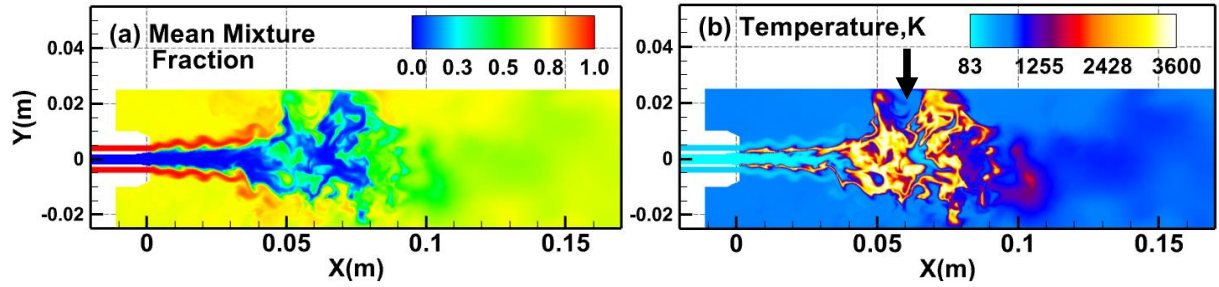
**FIG.5.** Instantaneous contours along the axial plane of the chamber: (a) Mixture fraction, (b) Temperature

**B. Flame Features and Comparison**

This section provides a qualitative view of the flow and flame features of the validation case before comparison with experimental data. Figure 5(a) shows the mixture fraction contour at the axial plane, which shows higher intensity in the methane stream on top of the central oxygen stream. It also depicts the consumption of the central oxygen stream at an axial distance of 12D (diameter of LOx post) from the injector exit. The oxygen potential core length eventually decides the flame location. Figure 5(b) showcases the instantaneous turbulent flame temperature in the domain. It shows flame anchored at the LOx post location, which evolves from a shear layer into a radially expanded structure. The high-velocity gradient in the shear layer exerts strain on the flame and perturbs the flame surface with small-scale turbulent structures. The axial cut section shows a wrinkled, highly turbulent flame structure. It exhibits a cylindrical shape with low radial expansion till 25-30mm from the injector exit, followed by larger radial expansion till the end of the high-temperature region. The radial expansion occurs due to the boiling-like phase change phenomenon (pseudo-boiling[49]) of the oxygen stream. The instantaneous temperature contour indicates flame ending location close to 60mm from the injector faceplate. A high-temperature eddy is seen to separate from the main flame at the location marked in Figure 5(b).

A comparison of thermophysical properties in current LES and analytical models depicts the accuracy of the numerical framework. Figures 6(a) and 6(b) show the comparison of LES results (instantaneous) plotted at an axial cut (2D) plane in the Mascotte chamber with NIST[50] (National Institute of Standards and Technology) data. Figure 6(a) compares instantaneous specific heat in LES and NIST data, while Figure 6(b) displays a comparison for density. It shows that the thermodynamic model invoked in the LES framework can capture the abrupt change in specific heat and density close to the critical point of oxygen, as plotted in NIST. The abrupt change in LES data close to 150K is attributed to the transition of the LOx stream from a transcritical to a supercritical state. The SRK[43] EOS used in this framework closely captures the non-linear variation in specific heat and accurately represents the transcritical LOx injection density and a rapid decrease of density with temperature. It overall verifies the accuracy of the SRK thermodynamic model in the current LES framework to capture non-linear variation close to the critical point or, in general, the transition of the transcritical LOx stream to the supercritical state. The previous study[51] in our research group also displays the SRK model to accurately represent the thermophysical properties over the entire range of pressure-temperature variation.

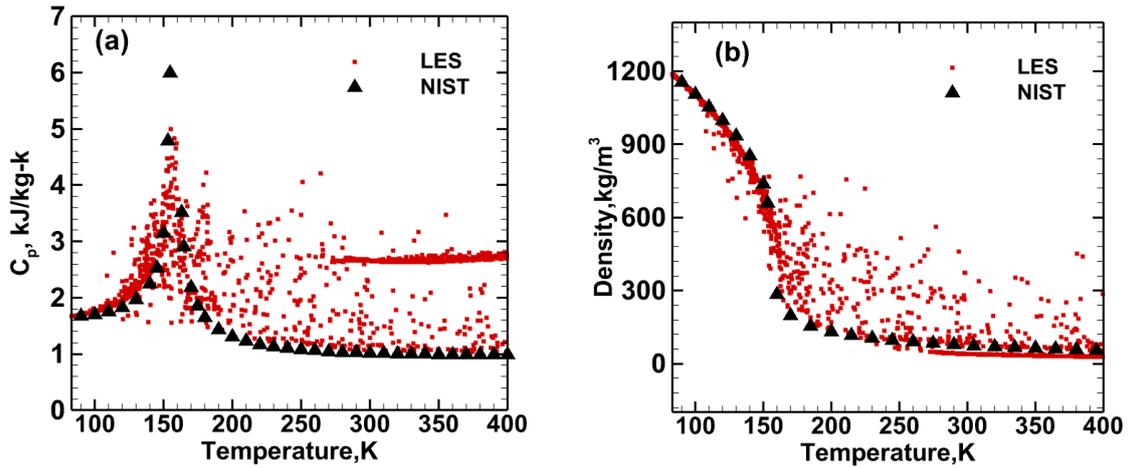

**FIG.6.** (a) Comparison of Specific heat (b) Density variation in LES with NIST data

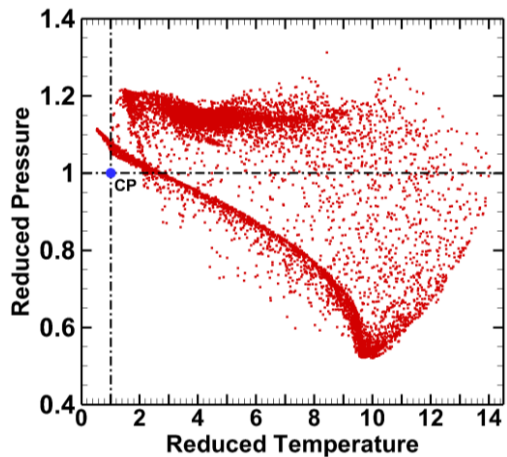

**FIG.7.** LES flame states in $P_r$-$T_r$ space

Figure 7 provides all LES flame states plotted at an axial plane for the reduced pressure and temperature. It highlights the absence of subcritical multiphase condition with no flame state below unity reduced pressure and temperature coexistence point (CP). The reduced pressure decreases with an increase in a reduced temperature, which can be attributed to the high critical point $H_2O$ product in the flame zone. The LES confirms no phase separation at the Mascotte chamber (G2) operating pressure. It should be noted that any further increase in operating pressure will move the flame states further away from the coexistence point (CP), eliminating any possibility of subcritical phase change.

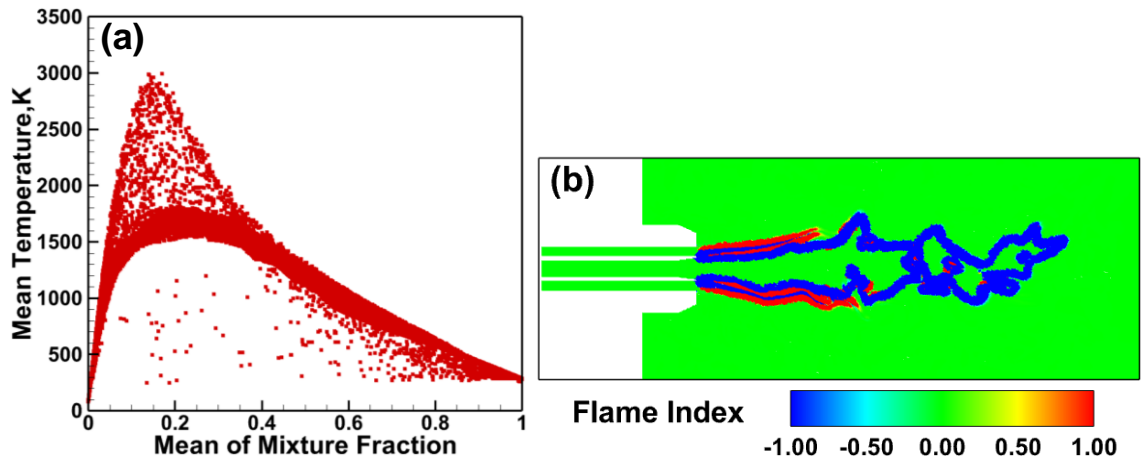

**FIG.8.** (a) Mean temperature distribution, (b) Flame index at the axial plane

Figures 8(a) and 8(b) illustrate the role of premixing in high-pressure LOx-methane flames. Figure 8(a) displays mean temperature data at an axial plane in mixture fraction space. It shows a mean temperature peak close to the stoichiometric mixture fraction 0.2. The graph shows the nature of non-equilibrium/kinetic effects built into the LOx-methane chemistry, which is dealt with appropriately in the FGM framework. Figure 8(b) displays the instantaneous Flame Index contour, which is defined as $FI = \nabla Y_{O_2} \cdot \nabla Y_{CH_4}$. Yamashita et al.[52] proposed that the FI distinguishes between premixed and diffusion flame regimes. The positive value of the flame index represents a premixed flame, while the negative value signifies a diffusion flame. Fuel and oxygen diffuse in the same direction in a premixed flame, and the index is positive; however, in a diffusion flame, fuel and oxygen diffuse from opposite directions, and the dot product is negative. Figure 8(b) shows a positive value in the injector outlet region, which indicates strong premixing and finite rate effects in the near-field injector flame. The negative value in the downstream region describes a predominantly diffusion nature of the flame. The FGM formulation represents the partially premixed nature of the flame with contribution from finite rate kinetics and turbulent diffusion appropriately.

Figure 9 compares OH concentration in the LES and G2 test cases. The emission imaging of excited OH radical was recorded in the G2 test and time-averaged to find the mean flame structure. Only a qualitative picture of OH concentration is available from the test. The bottom part of Figure 9 displays the Abel-transformed OH concentration image in the G2 test, whereas the sampled OH concentration from LES is shown at the top. The sampled LES contour represents the ensemble average of instantaneous OH mass fraction over a considerable flow time, with the sample acquired at every 0.1 ms. LES captures flame to stabilize on the injector lip, as seen in the G2 test snapshot. A higher temperature region (red color) in LES and light emission in the G2 test are seen to emanate from a cylindrical envelope in the shear layer region. A rapid radial expansion is noticed in the LES and G2 test after an axial location of 0.04m. The central (blue) region corresponds to the oxygen jet, and the flame engulfs the oxygen core till the end. The current LES shows a favorable comparison of OH concentration with experimental observation. A good match in the flame length and radial expansion is observed in the LES and test. In both cases, the flame terminates at an axial location of 60mm from the injector exit. Although some differences in profile are noticed towards the end of the flame, LES captures the flame anchoring, shear-layer burning, transcritical oxygen pseudo-boiling led radial expansion, and overall flame potential core in a close match with benchmark test, which fulfills the primary goal of the validation study.

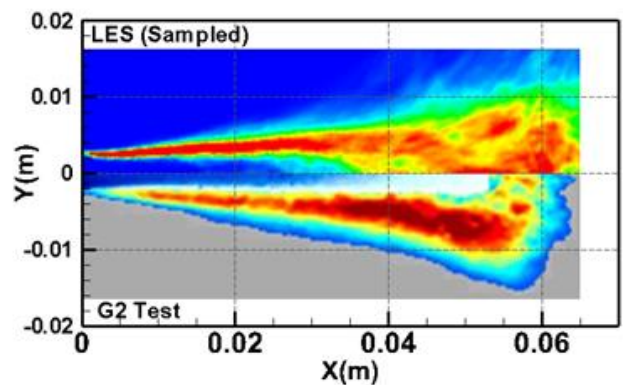

**FIG.9.** OH concentration comparison with G2 test case

The validation study reveals the utilized numerical framework's efficacy in capturing transcritical injection and supercritical LOx-methane combustion. It showcases a single-phase modeling strategy to accurately describe the high-pressure LOx-methane flame. It shows turbulent mass transfer/mixing as the most influential modeling process above the critical point operating condition. The validated numerical method is now applied to the multi-injector element to decipher the underlying LOx-methane supercritical combustion dynamics in a designed swirl coaxial injector combustor in the next section.

## IV. RESULTS AND DISCUSSION

A multi-element combustor's combustion instability mechanism is complex due to injector-to-injector and injector-chamber flow and flame interactions. This study utilizes multiple bi-directional swirl coaxial injector elements, which are preferred for large flow rates, higher thrust per element, and higher combustion efficiency. This section presents the impact of multiple swirl injectors on LOx-methane flow and flame dynamics, the role of feedback coupling between the chamber and injectors, and the effect of lower fuel injection temperature on combustor stability. The study highlights flow physics inside the injector, the role of key injector design parameters, and the mechanism responsible for pressure fluctuations. The dynamic response of bi-directional swirl injectors is revealed, which can govern the overall performance and stability of the combustor.

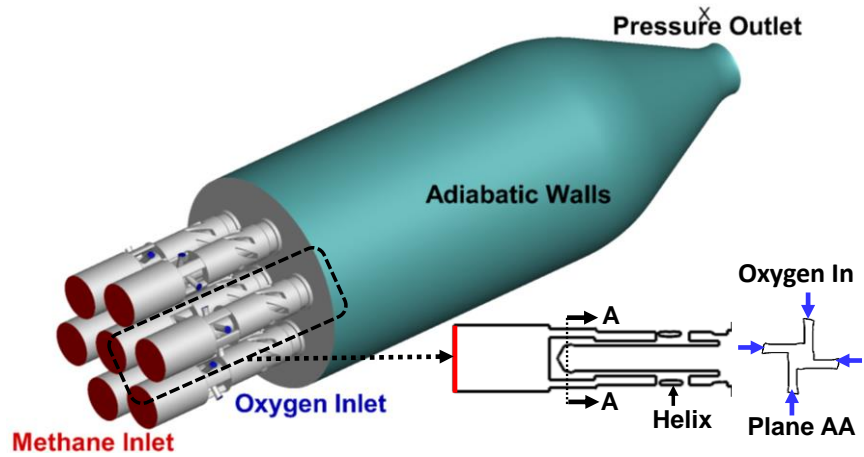

**FIG.10.** Computational domain- Multi-Element Combustor

### A. Multi-Element Model Configuration and Boundary Conditions

We have considered a multi-element combustor designed to accommodate 7 bi-directional swirl injectors. It is comparable to a full-scale rocket engine and can provide a relevant understanding of multi-element injection, mixing, and combustion dynamics. The designed combustor has an axial length of 13 times the throat diameter and a chamber diameter of 4.18 times the throat diameter. The injector plate accommodates seven elements, with one injector at the center and the other six placed hexagonally and 17.5 mm from the center. Figure 10 illustrates the computational domain with seven swirl coaxial injectors truncated at the throat location. A separate cut view of the injector depicts methane and oxygen's entry and flow path. The propellants enter the combustor as separate non-premixed streams and exhibit rapid turbulent mixing at the injector exit region. The injector elements are bi-directional swirl coaxial, with oxygen flow through the center path and methane through an annular flow path around oxygen. The injector configuration is derived from the propulsion system successfully operating on LOx-hydrogen propellants. A bi-directional swirl is provided by tangential entry to the oxygen and helical vanes in the axial passage of methane to acquire swirl motion. The oxygen path axial length is $11.3R_n$, and the coaxial methane flow path length is $12.25R_n$, where $R_n$ is the oxygen post radius. A positive recess of $0.6R_n$ is provided to the oxygen post to enhance mixing. The bi-directional swirl creates overlapping oxygen and methane flow cones, leading to rapid mixing at the shear layer along the axial direction. In this study, LOx-methane injection conditions are close to exact rocket operating conditions, with the flow rate per injector element fixed to maintain chamber pressure close to 64 bar. Oxygen is injected in the transcritical state with an injection pressure of 70 bar and injection temperature of 83K, while methane is injected in the supercritical state with an injection pressure of 70 bar and at an injection temperature of 230K. The corresponding injection density is 88 and 1190 kg/m$^3$ for methane and oxygen, respectively. A significant difference in sound speed is noticed in methane (350.25 m/s) and oxygen (990.33 m/s) streams, which can impact the interaction of chamber acoustic waves with upstream injectors. The injection conditions are similar to the Mascotte validation study but operate close to stoichiometry with an overall oxygen to fuel mass ratio (O/F) of 3.4, corresponding to an equivalence ratio of 1.17.

Mass flow inlet boundary condition is imposed at methane and oxygen injector inlets. Walls are treated as an adiabatic, and no-slip condition is imposed. The domain is truncated at the throat, automatically providing acoustically fully reflective boundary conditions due to the choked throat condition. A unity Mach number at the throat is ensured to maintain acoustically reflective boundary conditions in the simulation. In this study, two refined grids: fine with 4.5 million cells and a finer mesh of 6.5 million cells, are tested. A higher number of cells are packed close to the injector outlet region, with minimum cell sizes of 0.35mm and 0.22mm in fine and finer mesh, respectively. Figure 11 compares the mean temperature along the domain axis for two different grid sizes, i.e. fine and finer mesh. It indicates a minimal temperature difference between the two grids and displays a consistent trend along the axis. As section III (validation) mentioned, the LES resolution criterion is also used to determine grid size, which can resolve more than 80% of turbulent scales at comparatively lower computational expense.

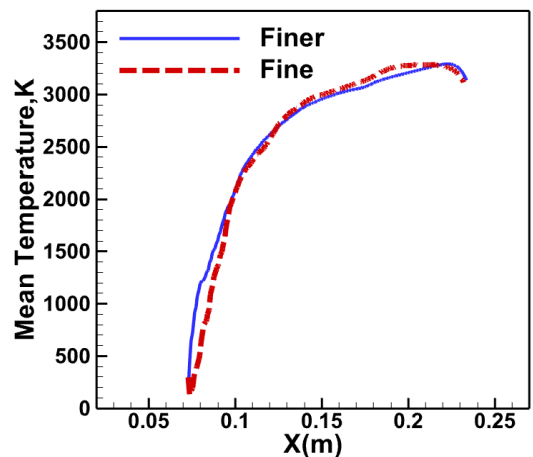

**FIG.11.** Axial temperature variation for two grids

Given the similarity in temperature variation between the fine and finer grids and to reduce the runtime of multiple LES cases, the grid with the lower mesh count is used for all subsequent analyses. The validated LES methodology is applied for this case, with spatial and temporal discretization performed using the second-order and bounded second-order implicit method, respectively. The acoustic CFL number is defined based on the flame region's acoustic velocity and mesh size. Our earlier CVRC study[16] showed the necessity to keep a lower acoustic CFL(ACFL) value to capture pressure oscillations and minimize acoustic wave dissipation. It showed that an ACFL value of 5 or below captures wave dynamics. In this study, the ACFL/time step size is fixed in such a way as to capture the unsteady injector flow and combustion dynamics without significant numerical dissipation and in a reasonable computational time. LES is performed for adequate flow-through times to ensure an initial condition independent solution and to collect sufficient samples of unsteady statistics. We have performed LES for numerous acoustic cycles to capture the unsteady physics in this multi-element injector and chamber region. Initial analyses include the flow and flame dynamics in the combustor through instantaneous contours followed by detailed spectral analysis of probe point data. The evolution of dominant frequency modes in the combustor and injector region reveals the mechanism of combustion dynamics. Spectral analysis of chamber probe data demonstrates the effect of lower fuel injection temperature on the stability of the combustor.

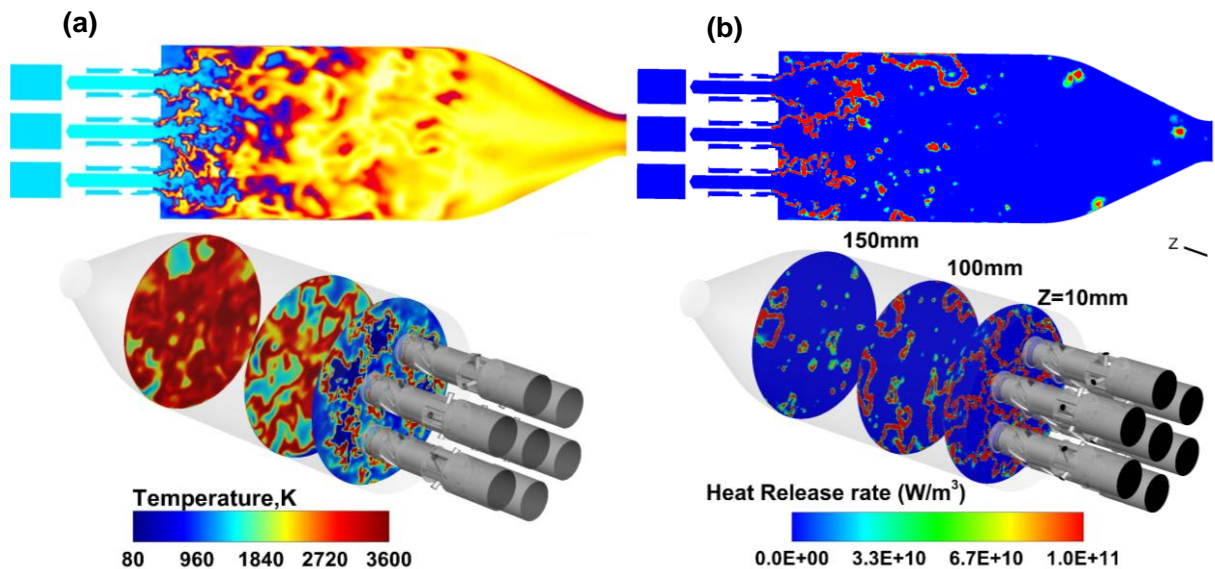

**FIG.12.** (a) Instantaneous temperature, (b) heat release variation at axial and radial cut plane

**B. Instantaneous Flow Fields**

The instantaneous contour plots at a particular time instant show significant flow and flame features. Figure 12(a) shows the instantaneous flame temperature contour at the axial (top) and radial (bottom) cut plane. It displays the evolution of supercritical LOx-methane flames, with flame-flame and flame-wall interaction. It displays a flame anchored at the end of the oxygen post in the recess region, exhibiting high temperature in the shear layer between the oxidizer and fuel close to the injector exit. The higher temperature seen in the shear layer region corresponds to stoichiometric conditions. A corrugated flame structure is visible in the shear layer at the Z=10mm radial plane, further evolving into intense combustion and homogenous flame mixing in the injector downstream region. The high-temperature turbulent eddies break apart from individual injector flames and merge into each other, forming a well distributed high-temperature zone.

Figure 12 (b) presents the instantaneous heat release rate variation at the axial and radial cut plane. An intense heat release rate is seen in the injector downstream region, particularly at the shear layer location, which indicates the diffusion-dominated nature of the LOx-methane flame. The radial variation also presents a highly wrinkled heat release rate, visible till the consumption of fuels, i.e. Z = 100mm. Figure 13 shows the contour of the instantaneous flame index (FI) at the axial cut plane. The contour at the injector near-field region highlights the diffusion-dominated nature of the flame, with a negative value of the flame index in the shear layer and the downstream region, indicating a non-premixed mode of combustion.

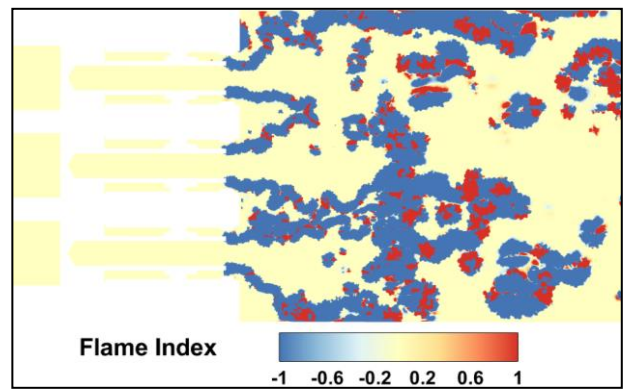

**FIG.13.** Flame index

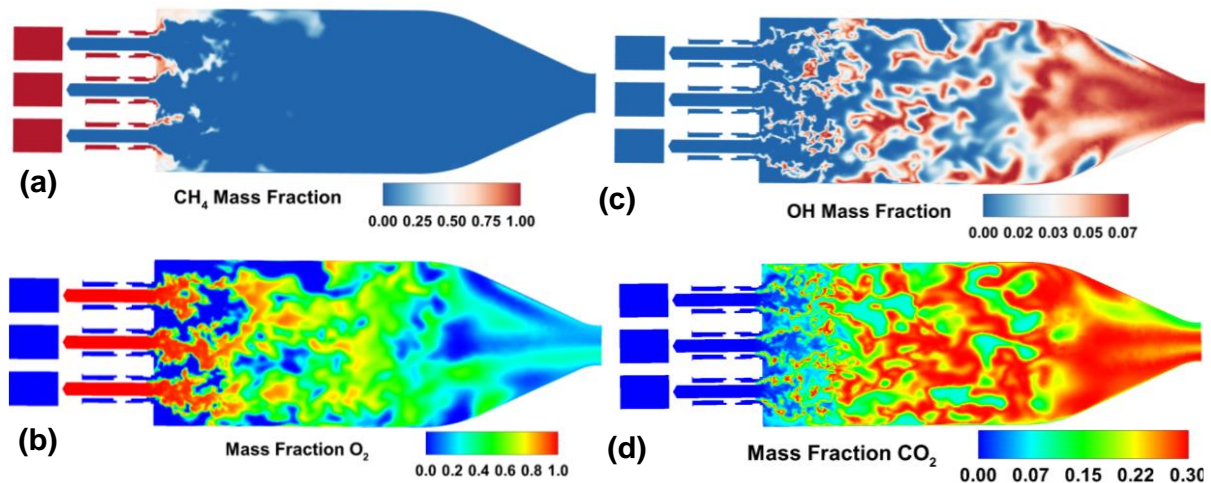

**FIG.14.** Instantaneous Mass Fraction contours: (a) $CH_4$ (b) $O_2$ (c) OH (d) $CO_2$

The positive value of the flame index representing premixed flame is also present in small pockets over the widely present diffusion flame structures. The FI contour shows the presence of non-premixed and premixed combustion modes in this multi-element configuration, establishing the requirement of the FGM model. Figure 14 depicts the instantaneous flow features, showcasing methane, oxygen, OH, and $CO_2$ mass fractions along the axial length of the combustor. Figure 14(a) shows a high methane mass fraction in the injector region, which dissipates along the swirl cone in the combustor. A higher oxygen mass fraction is visible in the oxygen post and close to the injector exit, displayed in Figure 14(b). The oxygen mass fraction decreases along the axis, with dense oxygen consumption occurring within the cylindrical section of the combustor. Figure 14 (c & d) displays OH and $CO_2$ mass fractions at the axial plane. Both OH and $CO_2$ mass concentration initiates from the heat release zone in the shear layer, with higher concentration in the downstream region indicating overall combustion within the cylindrical section of the combustor. The instantaneous contours reveal physical flow and flame characteristics, emphasizing the numerical approach established in the validation study for effectively modeling supercritical combustion within the multi-element combustor configuration.

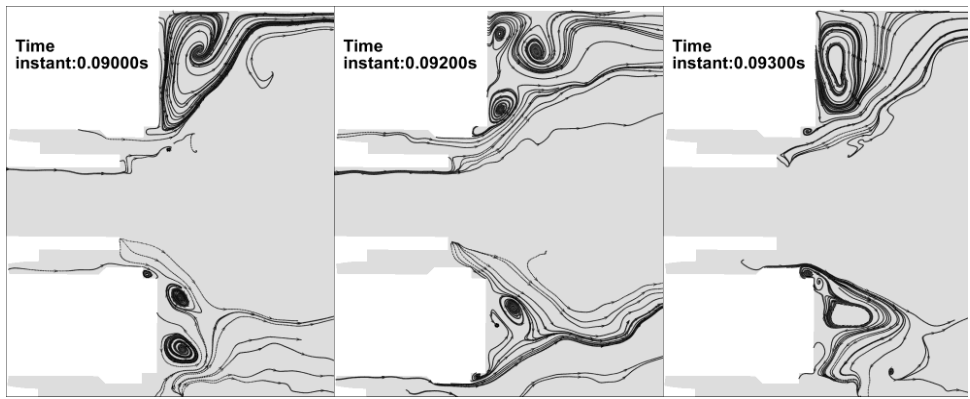

**FIG.15.** Temporal evolution of streamlines at injector near field axial plane

**C. Swirl Injector Flame Dynamics**

The swirl injectors provide rapid mixing and help in flame stabilization. The bi-directional swirl motion of methane and oxygen generates toroidal recirculation zones. In this configuration, methane and oxygen form clockwise and anti-clockwise overlapping swirl cones, which leads to strong shearing and enhanced mixing close to the injector exit. Figure 15 shows the temporal variation of streamlines from the injector close to the wall. It depicts a big recirculation zone close to the wall and two counter-rotating bubbles between the injectors. The recirculation zones are fuel-rich and contain hot products, which provide a necessary flame-holding mechanism. The streamline contour displays the formation/separation of bigger recirculation bubbles from or into multiple smaller bubbles, respectively. These corner recirculation zones continuously provide hot products and keep the flame anchored in the recess region. It maintains intensive burning in the shear layer and the region downstream of injectors. The flame is swirl-stabilized, forming a vortex core at the injector exit. The vortex core is formed due to the centrifugal force generated by the swirling motion of methane and oxygen. The injector swirl dynamics are further illustrated through vorticity magnitude. Figure 16 highlights the complex vorticity dynamics in this multi-element computational domain. It reveals several distinct features associated with flow in injectors and the combustor region. It presents the distribution of vorticity magnitude at an axial cut plane and radial planes at different axial locations within and outside the injectors.

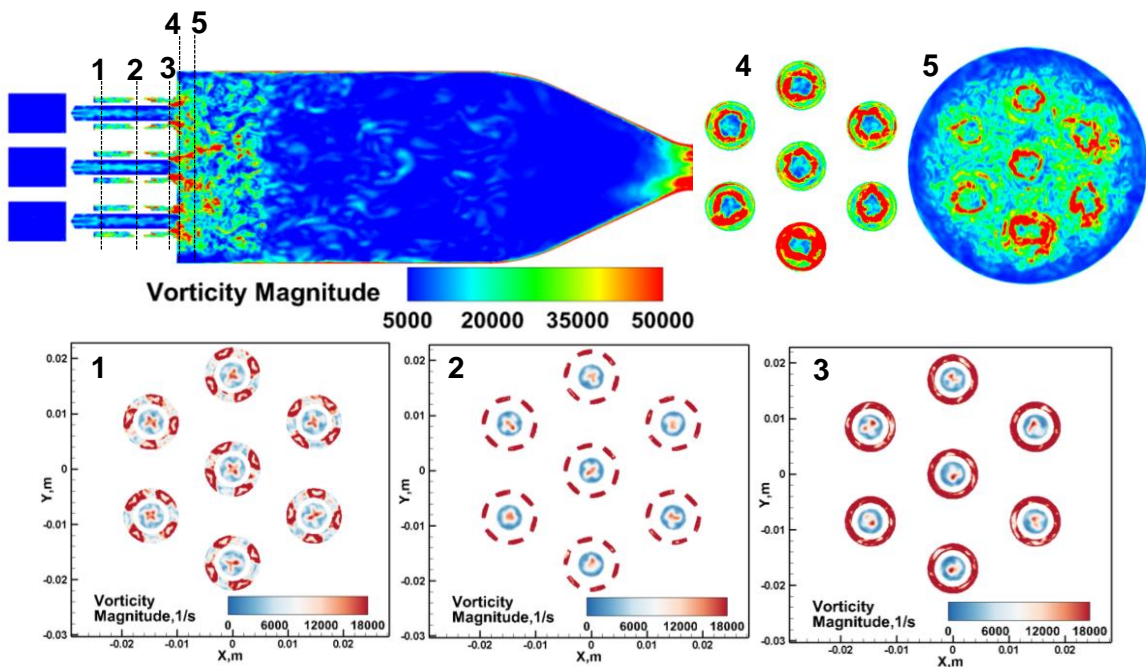

**FIG.16.** Instantaneous vorticity magnitude at axial and radial planes

The vorticity magnitude at the axial cut plane shows the presence of a strong vorticity zone emerging from the recess location. The presence of flame results in large temperature/density and velocity gradients in the recess region, which eventually leads to higher vorticity magnitude in the recess and post-recess location. Planes labeled 1, 2, and 3 display vorticity features within the injector and 4 and 5 at the injector exit and combustor, respectively. The injector planes 1, 2, and 3 show lower vorticity magnitude in the LOx film adjacent to the oxygen injector walls. In contrast, a high vorticity region is seen at the central core at every axial location. The methane flow presents a high vorticity magnitude at all injector planes. Plane 4 highlights higher vorticity generation at the shear layer between the fuel and oxidizer. In contrast, Plane 5 illustrates the presence of vorticity at the shear layer and the distribution of vorticity over a large number of eddies. It is well known that the vorticity in the near field injector region can influence the stability of combustion. The vortex breakdown can trigger the onset of enhanced combustion dynamics and is an important aspect associated with bi-directional swirl injectors. The flow dynamics can lead to enhanced pressure and temperature fluctuations. The effect of flow dynamics on the overall stability of the combustor is assessed in the next section.

### D. Spectral Analysis and Combustion Dynamics Mechanism

Combustion dynamics are assessed through the evolution of fluctuating pressure waves in the combustor. The temporal evolution of pressure can reveal the onset of any dynamic activity in the combustor. Figure 17(a) displays the radial distribution of instantaneous absolute pressure along the length of the combustor. It depicts the onset of dynamic activity with the transverse movement of a pressure wave in the combustor. The pressure contour exhibits a standing tangential acoustic wave in the combustor, with a typical pattern of maximum pressure near the combustor wall to a lower value at the chamber center. It highlights the clear formation of tangential mode pressure oscillation at a radial plane, Z=100mm. The radial distribution of pressure at a plane, Z=10mm, displays small zones of high and low pressure distributed all over the plane. The dominant features of this pressure wave are extracted through spectral analysis of pressure probe data. The time histories of pressure and other parameters are collected at different probe locations in the combustor and the injector's oxygen path. Figure 17(b) shows the probes placed in the injector and chamber region. The injector probes are denoted by symbol, IP, and chamber probes by CP. Multiple probes are placed to collect unsteady statistics and reveal the dynamics mechanism observed in this multi-element configuration. The simulation is performed for 40 flow-through times before collecting the pressure statistics to avoid the effect of initial transients. Absolute pressure is collected at a sampling frequency of 0.1MHz.

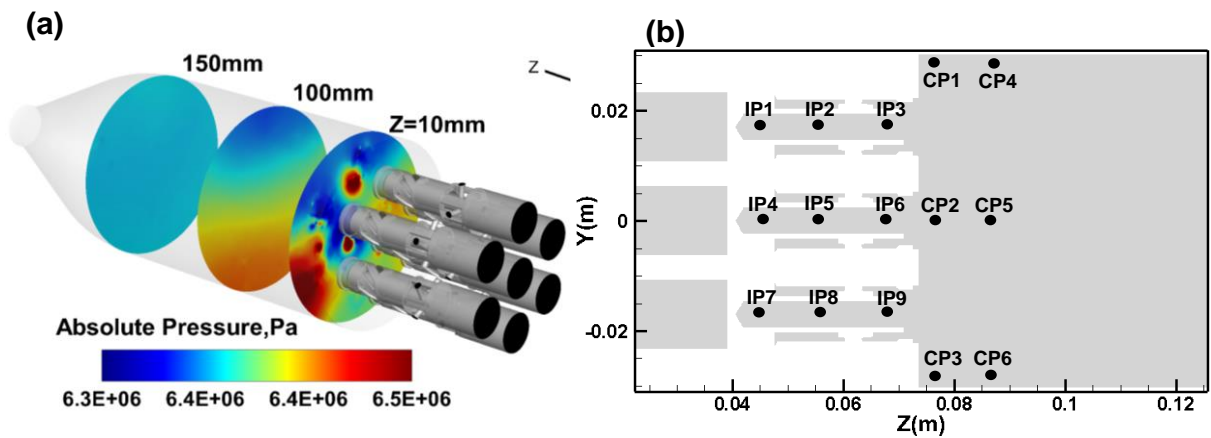

**FIG.17.** (a) Absolute pressure variation at different radial planes, (b) Probe location at an axial plane

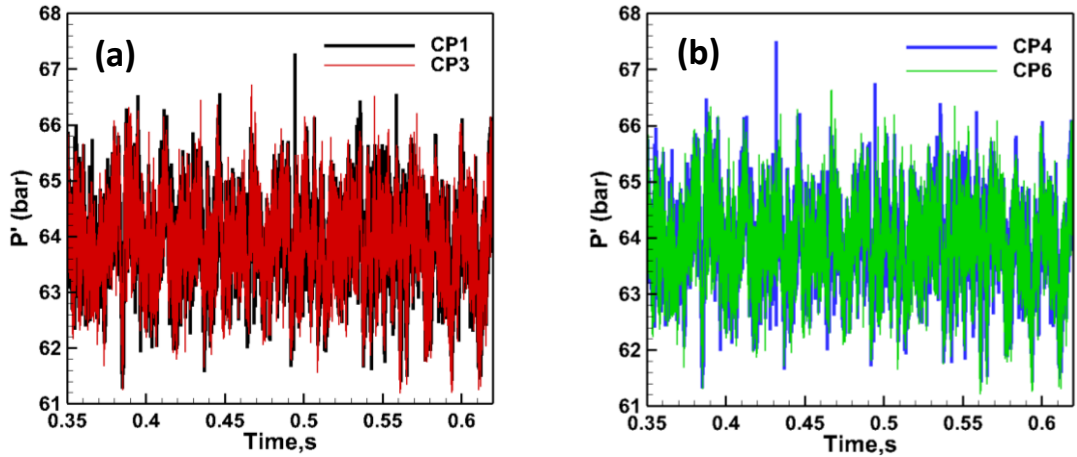

**FIG.18.** Pressure variations at near-wall chamber probes: (a) CP1 & CP3, (b) CP4 & CP6

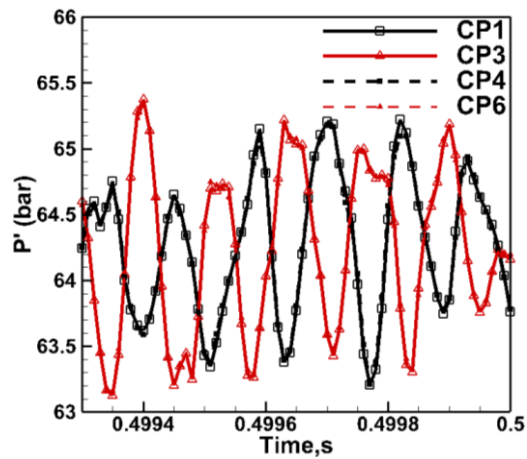

**FIG.19.** Zoomed-in view of pressure variations at near-wall chamber probes

Figures 18 (a) and (b) depict raw absolute pressure time histories of near-wall chamber probes (CP1, CP3, CP4, CP6). Figure 18(a) displays the peak-to-peak pressure variation above the mean pressure of 64 bar at chamber points CP1 and CP3. A similar pressure statistic is observed for chamber probes CP4 and CP6, as shown in Figure 18(b). Figure 19 reports a zoomed-in view of the pressure variation at all near-wall chamber probes in a time window of 0.7ms. The time window displays 5.5 cycles of pressure wave, corresponding to an acoustic frequency close to 8000 Hz. The diametrically opposite probe data reveals a pressure phase variation between channels CP1, CP3, CP4, and CP6. A 180º out-of-phase variation is captured in CP1-CP3 and the same as in CP4-CP6 data. There is no phase difference in CP1-CP4 and similarly in CP3-CP6 probe locations. The 180º phase difference in diametrically opposite near wall probes indicates the evolution of tangential standing wave in the combustor (Figure 17(a)). The raw pressure variation displays the presence of transverse acoustic waves, which is further illustrated through analytical and spectral analysis.

An analytical estimate on acoustic mode frequencies (*f*) using a mean sound speed value of 1200m/s, calculated from CEA (chemical equilibrium and applications) equilibrium solver at a chamber pressure of 64 bar and oxygen to fuel mass ratio (O/F) of 3.4. The resonant frequencies are calculated via:

$$f_{m,n,q} = \frac{c}{2\pi}\sqrt{\frac{\beta_{m,n}^2}{R_c^2} + \frac{q^2\pi^2}{L_c^2}} \quad (4)$$

where c denotes the speed of sound in the chamber, $R_c$ and $L_c$ represent the radius and length of the combustor, $\beta_{m,n}$ are roots of the Bessel function, and *m, n*, and *q* are the mode numbers. The theoretical dominant frequency of the combustor at the first longitudinal (IL) mode is 4347Hz, and the first tangential mode (1T) frequency is 11630 Hz. Figure 20 displays the frequency spectrum (Fast Fourier Transform-FFT) plot of all near-wall combustor probes, showing two peaks at 3.5 to 4.05 kHz, respectively, and a sharp frequency tone at 8.02 kHz. All near-wall combustor probes depict the same frequency peaks. The frequency peak at 3.5 kHz is attributed to hydrodynamic instability associated with vortex shedding. The vortex shedding frequency calculated at Strouhal number of 0.2 is 3450 Hz, which is very close to the first peak observed in Figure 20. The second and third peak close to 4 kHz and 8 kHz is identified by comparing the FFT spectrum of a near wall (CP1) and central (CP2) probe. However, the pressure variation of near-wall probes (Figure 20) clearly demonstrates the onset of transverse acoustic waves at 8 kHz.

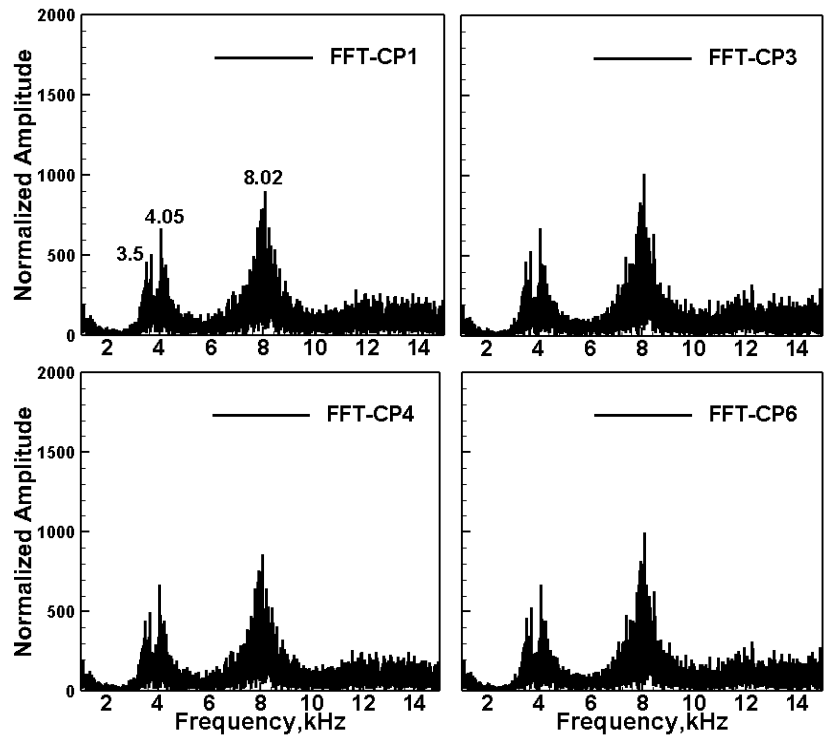

**FIG.20.** FFT of near-wall chamber probes

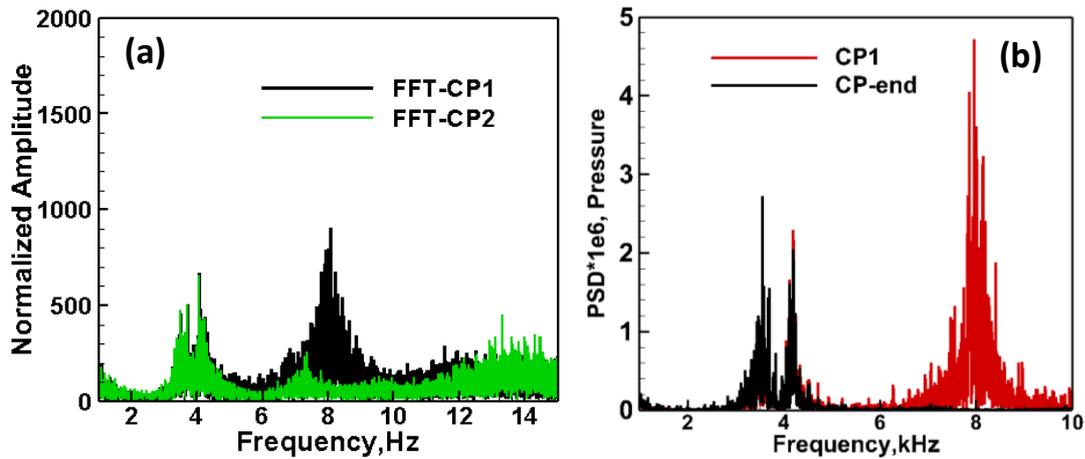

**FIG.21.** FFT comparison of the near wall probe with central and end probe: (a) CP1 & CP2, (b) CP1 & CP-end

Figure 21(a) compares the frequency spectra of CP1 and CP2. The central combustor probe (CP2) shows the same 3.5 kHz and 4 kHz frequency except for a sharp peak at 8 kHz as in the CP1 plot. The absence of an 8 kHz frequency peak at the center location confirms the presence of the first tangential mode (1T) near the combustor walls. The 4 kHz frequency observed in the FFT spectrum is considered the first longitudinal (1L) resonant mode, close to the analytical value calculated at constant equilibrium sound speed. Figure 21(b) depicts the FFT spectra comparison of CP1 (chamber start) to a probe placed at the start of the nozzle convergent, CP-end. It shows the absence of 1T mode frequency at the chamber end probe but captures longitudinal mode frequency as seen in the CP1 probe. The two probe locations, i.e., chamber start (CP1) and at the start of the converging section of the nozzle (CP-end), represent two anti-nodal points of the combustor.

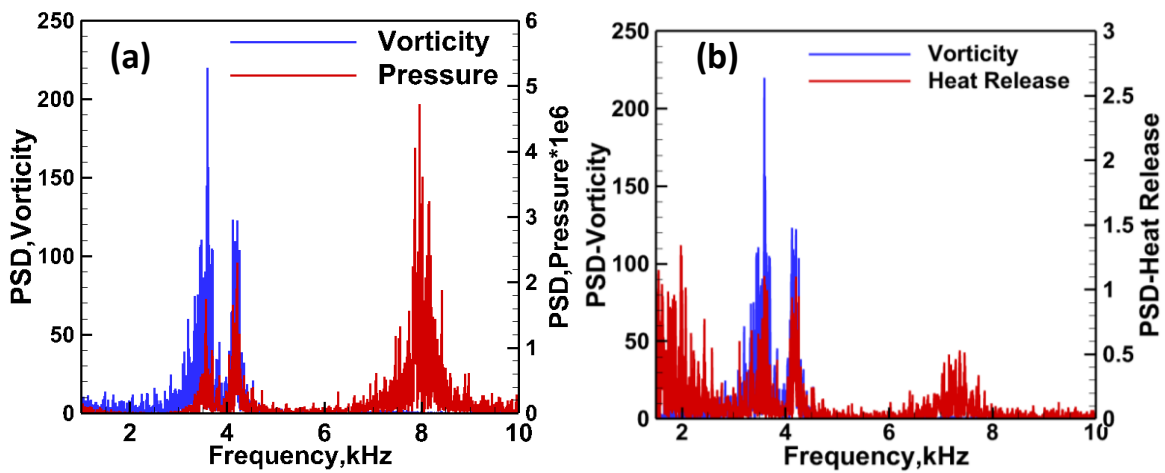

**FIG.22.** FFT comparison of the near wall with combustor central and end probe: (a) Velocity & Pressure, (b) Vorticity & Heat Release

Figure 22 (a, b) shows the FFT spectra comparison of vorticity with pressure and vorticity with heat release rate, respectively. The vorticity and pressure spectra display frequency match close to 4 kHz. The vorticity heat release frequency spectra also compare similarly (Figure 22 (b)). A close match of vorticity and heat release shows a coupling between hydrodynamics and combustion. The vortical structures can affect heat release distribution due to frequency coupling between vorticity and heat release, eventually leading to higher pressure fluctuations/enhanced dynamics in the combustor.

The spectral analysis identified combustor resonant modes (1L and 1T) at 4 and 8 kHz, respectively, whereas the analytical calculation based on mean sound speed provides a considerably higher value of 11630 Hz for the 1T mode. The local variation of sound speed in the combustor can impact the evolution of resonant modes, especially the transverse modes present close to the chamber wall in the vicinity of injectors. Figure 23 illustrates the time-averaged sound speed at the axial plane of the combustor. Simulation shows a considerable variation in sound speed in the combustor, which eventually affects the evolution of resonant frequency modes. It displays a lower sound speed near injectors than the combustor downstream region. It shows a sound speed of 820m/s in the injector near the field region close to the combustor walls. The analytical value of the first tangential mode calculated at the sound speed of 820 m/s leads to the first mode frequency of 8010 Hz, which is remarkably close to the peak frequency (1T mode: 8 kHz) observed in Figure 20. The central core region has a higher sound speed value due to homogeneous high-temperature gases. The first longitudinal acoustic frequency calculated at an average sound speed value of 1125 m/s is 4070 Hz, which matches the LES spectra frequency well. The modified frequency values calculated at local sound speed in the combustor closely match the frequency peaks observed in the LES spectral plots.

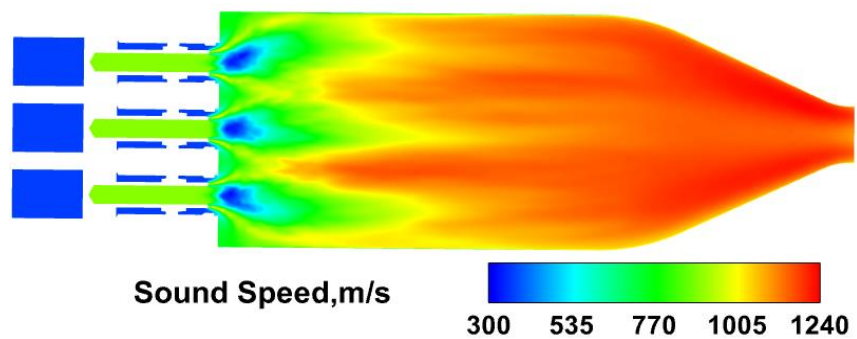

**FIG.23.** Time averaged sound speed at an axial plane

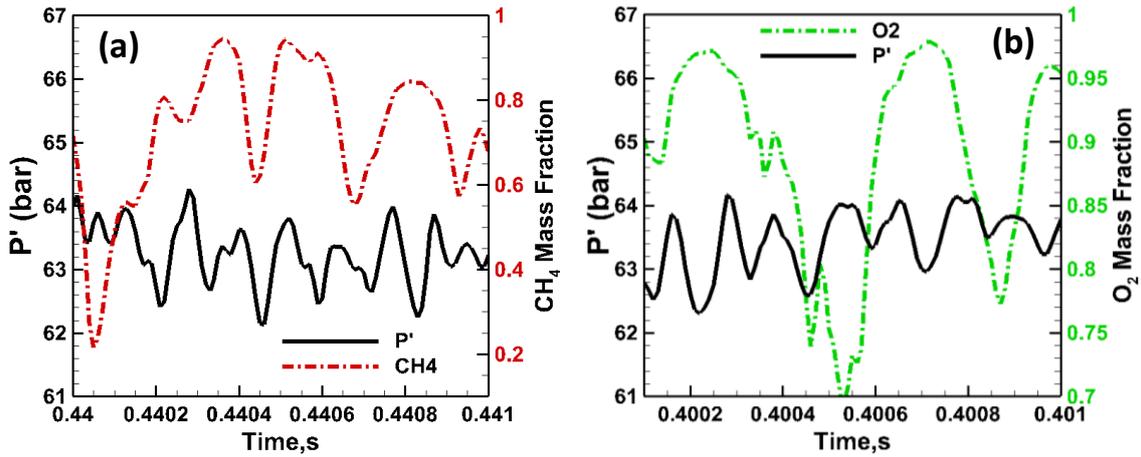
**FIG.24.** Temporal variation with pressure: (a) methane, (b) oxygen

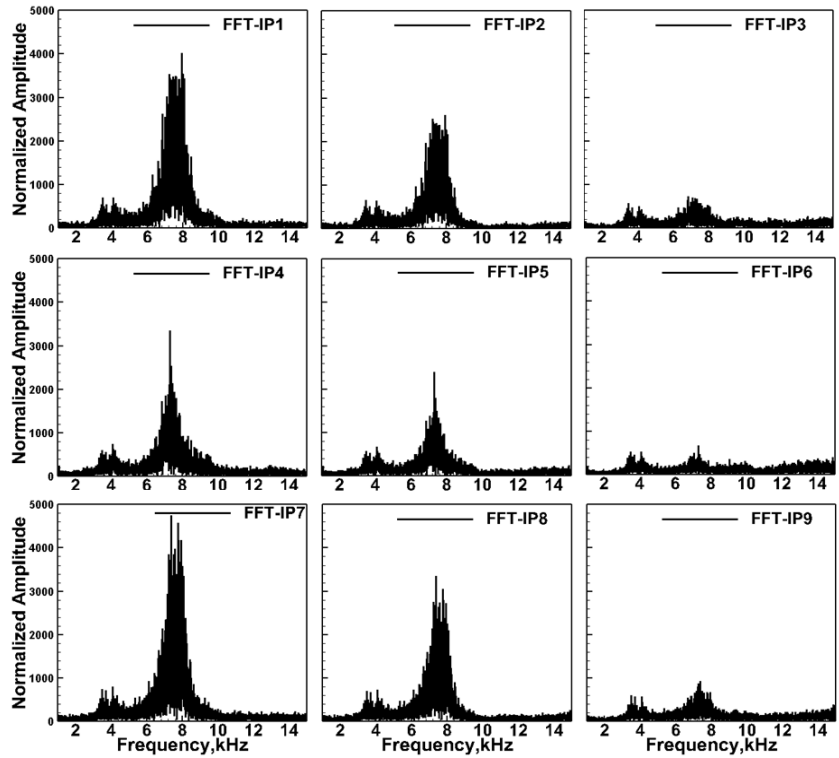
**FIG.25.** FFT of injector probes

### E. Effect of Injector Dynamics:

The onset of a transverse pressure wave in the combustor affects the injector flow rates and sustains the dynamics. Pressure probe data and spectral analysis are performed to reveal intricate injector flow dynamics. Any disturbance in the combustor can propagate into the injectors and induce pressure fluctuations, eventually leading to mass flow rate fluctuations. The methane and oxygen mass fraction fluctuations are observed in the near-wall injector. Figure 24(a & b) displays the temporal variation of methane and oxygen mass fraction along with instantaneous pressure. Probes are placed in methane and oxygen paths to collect the pressure and mass fraction statistics. It shows a higher mass fraction in the injector during the lower pressure zone in the combustor and vice versa. The mass fraction fluctuations can induce a mixture ratio (oxidizer/fuel mass ratio) variation in the combustor, leading to enhanced combustion dynamics, as observed in this case.

The pressure data in the injector region reveals the injector-associated combustion dynamics mechanism. FFT of probes placed in the injector section allows us to understand acoustic wave movement between the chamber and upstream injectors. The oxygen path of injectors can be acoustically treated as a quarter-wave resonator, with injector entry as a closed-end and exit as an open boundary[27]. It exhibits longitudinal modes with natural frequency calculated as $f = c/4(L + \Delta L)$, where L is the injector length, c is the speed of sound, and $\Delta L$ is the correction factor, $0.6R_n$. The theoretical first longitudinal mode (1L) frequency at an oxygen sound speed of 850 m/s is 7450 Hz. Figure 25 shows spectral graphs of each injector point with a peak frequency of about 7.5 kHz, which is very close to the theoretically predicted longitudinal mode frequency. Due to subtle changes in sound speed in the injector in both radial and axial directions, frequency peaks close to 7.5 kHz are seen in Figure 25, particularly in near-wall injector points. The sound speed in peripheral injectors is significantly influenced by the transverse wave in the combustor, resulting in a distributed frequency pattern close to 7.5 kHz, whereas a sharp tone is captured at central injector points (IP4, 5, 6). The presence of a 7.5 kHz frequency tone in central injector probes, the same as in peripheral injectors, confirms the longitudinal acoustic mode in injectors. Figure 25 also illustrates the quarter wave characteristics of the injector. The quarter wave representation is evident from the amplitude of peak frequency (7.5 kHz), which decreases from injector entry to the exit location. The amplitude drops from a maximum at the inlet probe IP1 to a minimum at the near exit probe, IP3, and similarly in the other two injectors. Longitudinal acoustic waves dominate the pressure oscillations in the injector, while the lower frequency modes near the injector exit are characterized by hydrodynamics like vortex shedding. The injectors at the periphery are affected by the onset of a tangential acoustic wave in the combustor. Figure 26 shows the raw pressure variation of injector probes IP1, 4, and 7. It displays a 180° out-of-phase variation for two opposite peripheral probes, IP1 and IP7, whereas the central element probe, IP4, shows a stable pressure trend. The out-of-phase pressure variation highlights the possible influence of combustor dynamics on the injector flow dynamics. The tangential wave in the combustor can impact the longitudinal acoustic modes in the injectors and vice-versa.

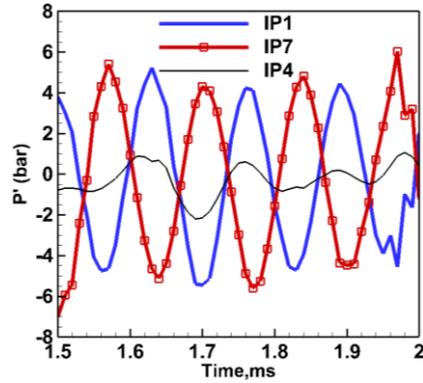

**FIG.26.** Pressure trace comparison of injector probes

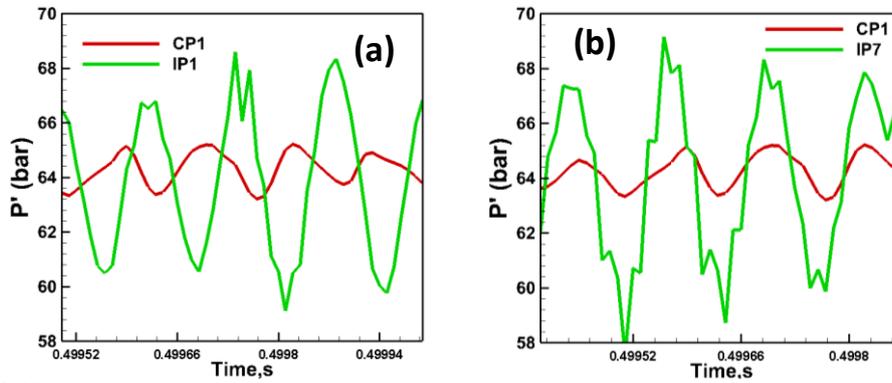

**FIG.27.** Pressure trace comparison of injector and combustor probes: (a) CP1 & IP1, (b) CP1 & IP7

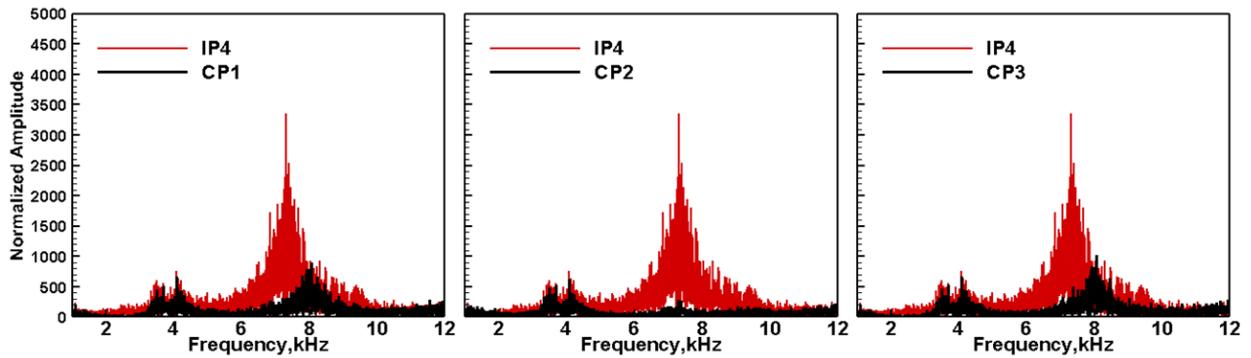

**FIG.28.** FFT spectrum comparison of injector and combustor probes

Pressure and dominant frequency explain the dynamics between injectors and combustors in this multi-element combustor. Figure 27 (a & b) highlights a pressure trace comparison between chamber probe CP1 and injector probe IP1, IP7. A large pressure oscillation in the injector probes IP1 and IP7 is noticed compared to the combustor probe CP1. The pressure trace displays an opposite phase trend in the CP1-IP1 plot (Figure 27(a)) and shows an in-phase variation in the CP1-IP7 plot (Figure 27(b)). The pressure trend shows coupled dynamics between the combustor and injectors, where a transverse acoustic wave in the combustor periodically influences pressure oscillations in the injectors, leading to mass flow rate fluctuations and resultant enhanced combustor dynamics. The pressure trace confirms the wave movement observed in the combustor and injector region (Figure 17(a)). Figure 28 compares the frequency between different combustor and injector probes, IP4. The injector probe IP4 is chosen for comparison, as tangential acoustic waves do not influence the central injector points. The frequency comparison shows injector-combustor frequency matches at 3.5kHz and 4 kHz, representing hydrodynamic and longitudinal combustor modes. Distinct peaks close to 8 kHz are seen in injector and combustor probes. As identified earlier, the tangential acoustic frequency of the combustor is 8 kHz, whereas the injector longitudinal acoustic mode is 7.5 kHz. A mismatch between the acoustic frequencies of the chamber and injector region is observed in this comparison. It also shows a distinct nature of acoustic modes present in the combustor and injector and the absence of frequency match in injector 1L and combustor 1T mode.

The above analysis demonstrates the mutual influence of chamber and injector acoustics. Similar frequency features in both the combustor and injector region are observed, which can result in a frequency-coupled system leading to enhanced combustor dynamics. The study showcases the influence of the combustor's transverse wave movement on the oxygen post's acoustic mode. Also, it shows that the oxygen path can act as a source of mass flow rate fluctuations and drive instability.

### F. Effect of Fuel Injection Temperature

This section presents the effect of lower fuel(methane) injection temperature on the stability of this combustor. Computation with a methane injection temperature of 200K is performed, and a detailed comparative analysis of flow and flame parameters for two fuel inlet temperature conditions, i.e. 233K (nominal) and 200K (off-nominal), is presented. Unsteady statistics are analyzed to determine the underlying mechanism responsible for the onset of high amplitude pressure fluctuations in low-temperature cases. Figure 29 displays probe locations in the chamber region. 12 probes are placed in the combustor region to capture unsteady pressure. Data is collected for over 80ms, corresponding to more than 300 cycles for 1L (first longitudinal) and 600 cycles for 1T (first transverse) mode, respectively.

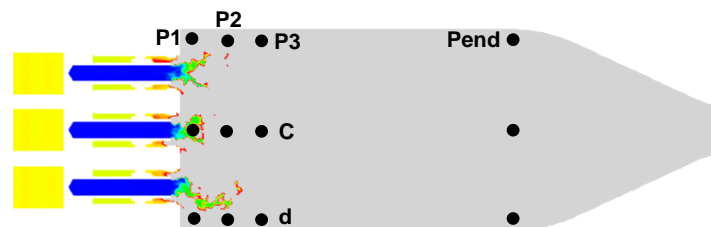

**FIG.29.** Probe locations at an axial plane. Label c refers to probe locations at the center. Label d refers to the probe location at the bottom. Label Pend refers to all probe locations at the end.

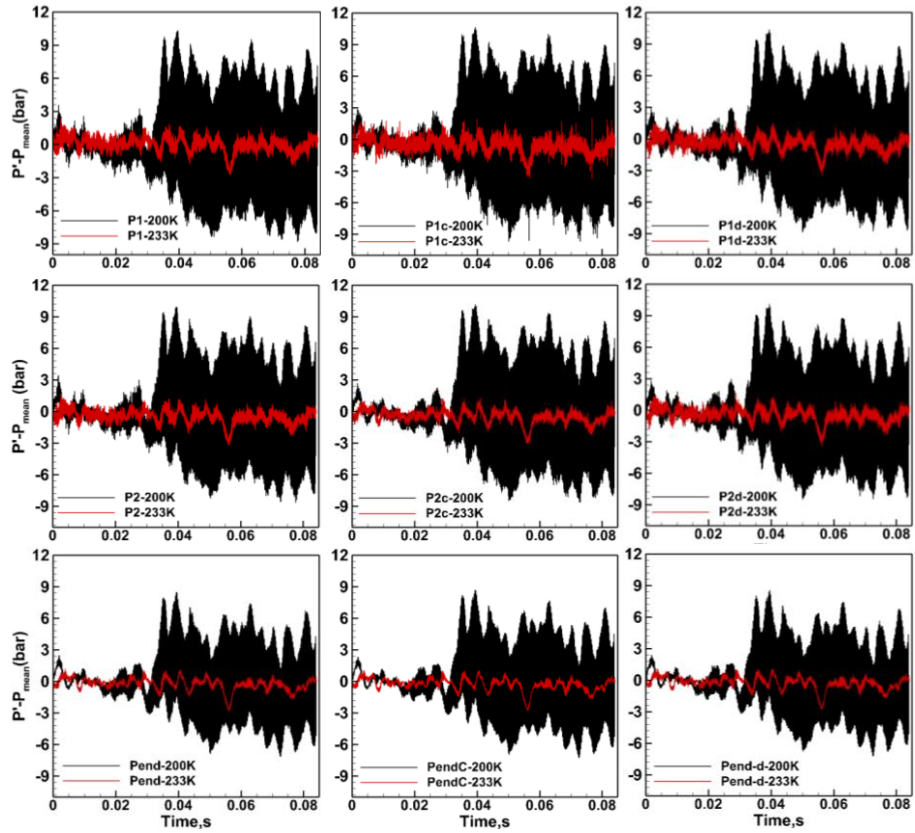

**FIG.30.** Unsteady pressure variation at probe location in the combustor

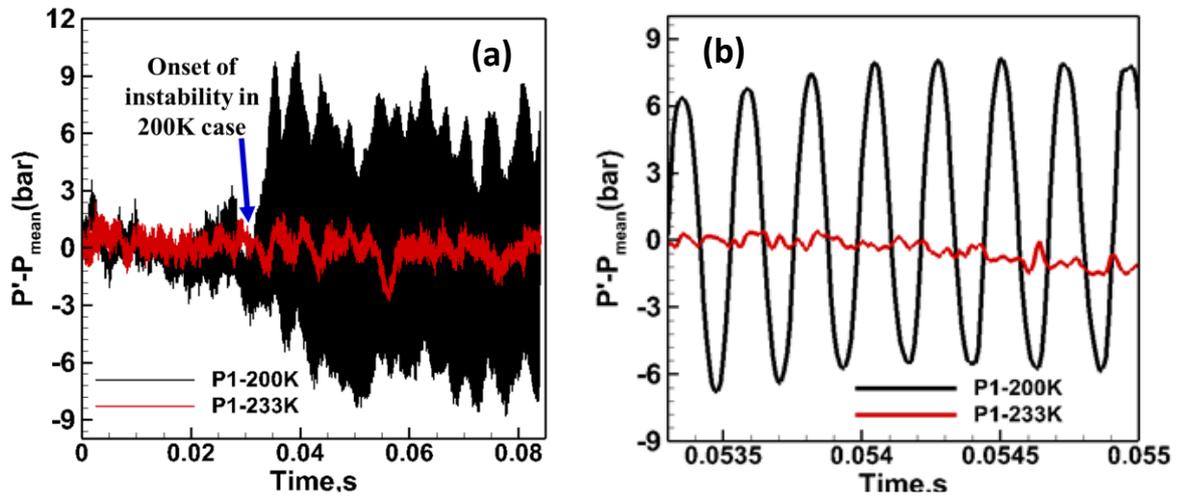

**FIG.31.** Pressure fluctuation comparison at P1 probe and peak-to-peak amplitude: (a) large time window, (b) small time window

Figure 30 compares fluctuating pressure in 200K and 233K cases at different probe locations. The unsteady variation reveals the presence of high amplitude pressure fluctuations in the 200K case at all probe locations, while the 233K case exhibits low amplitude. Figure 31 provides a separate view of pressure fluctuations at the P1 probe for both temperatures. The direct pressure trace comparison at the P1 probe in Figure 31(a) shows the onset of instability at a time instant of 0.031s for the 200K case. The violent pressure activity displays limit cycle oscillations and sustains till the end of the simulation. Figure 31(b) exhibits peak-to-peak amplitude close to 15 bar for the 200K case, while the 233K case displays extremely low amplitude pressure fluctuations. It displays the characteristics of a limit cycle with self-sustained pressure fluctuations.

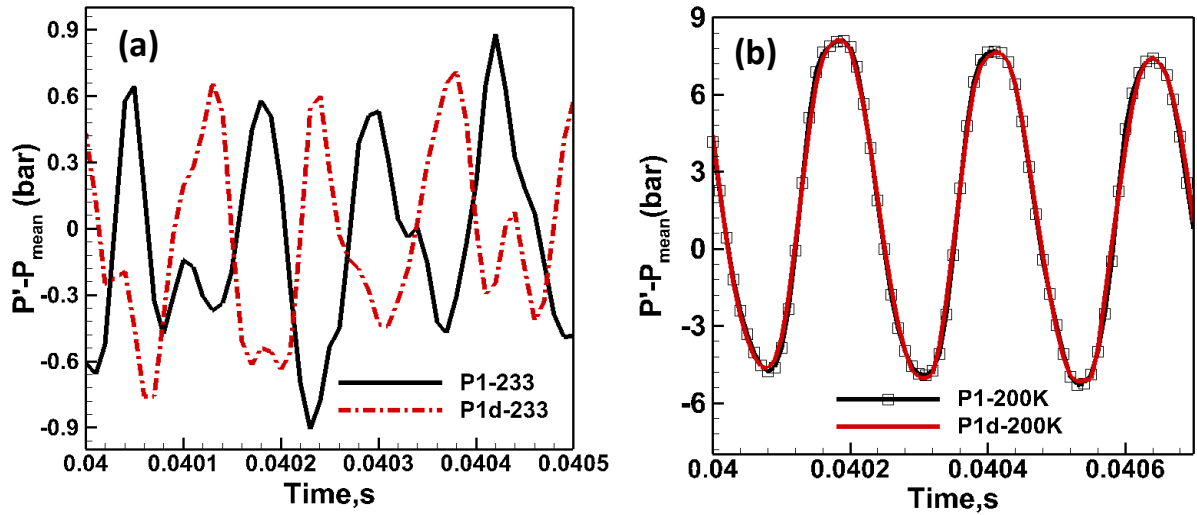

**FIG.32.** Pressure fluctuation comparison at P1-P1d probe location: (a) 233K, (b) 200K

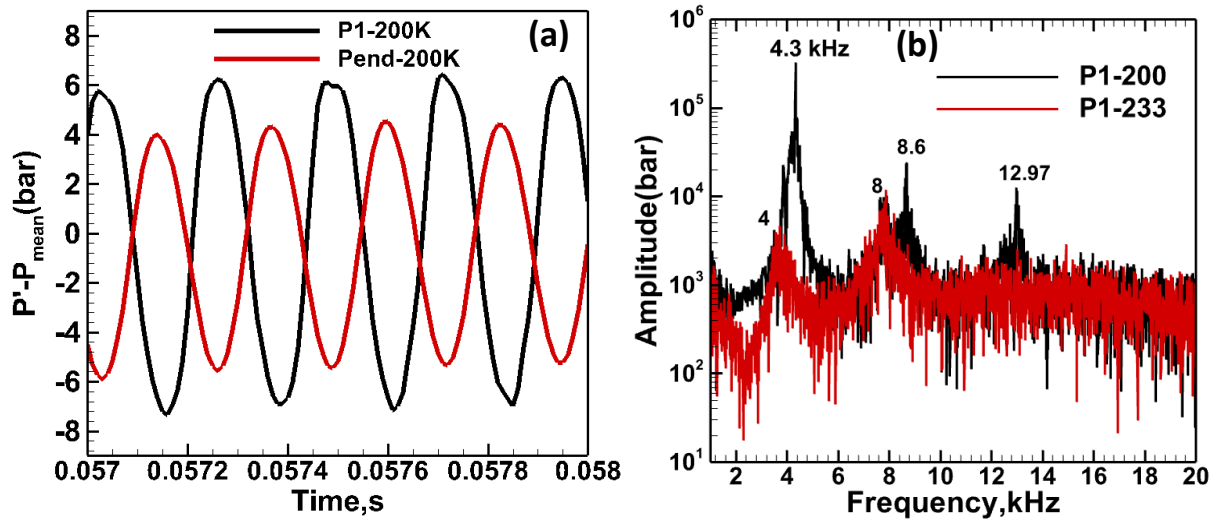

**FIG.33.** (a) Pressure fluctuation comparison at P1-Pend probe for 200K, (b) FFT comparison for 200K and 233K case

Figure 32(a & b) compares pressure fluctuations at probes P1 and P1d for 233K and 200K cases, respectively. Figure 32(a) displays low- amplitude pressure fluctuations for 233K case. An out-of-phase pressure variation at opposite-facing probe locations (P1, P1d) indicates a low amplitude transverse wave movement in the combustor. In contrast to the 233K case, Figure 32(b) shows a high amplitude, in-phase pressure wave for the 200K case, characteristic of longitudinal wave movement in the combustor. It shows the absence/extremely low amplitude transverse mode amplitude in such limit cycle oscillations. The pressure fluctuation trace in 200K further reveals the onset of wave movement in the combustor. Figure 33(a,b) displays pressure fluctuation variation at P1 and Pend probe locations. The out-of-phase pressure variation for the near injector (P1) and near throat (Pend) axial location also indicates the onset of a longitudinal acoustic wave in the combustor with a cycle frequency close to 4.3 kHz. Figure 33(b) compares the P1 probe frequency spectrum for both temperature cases, revealing the dominant frequency components associated with the longitudinal wave movement in the 200K case. It shows a distinct high amplitude peak of 4.3 kHz and its harmonics 8.6 kHz and 12.97 kHz in the 200K case, while such frequency tones are absent in the 233K case. The nominal case (233K) displays low amplitude 1L (4 kHz) and 1T mode (8 kHz), whereas the FFT spectrum of the 200K case shows high amplitude 1L, 2L, and 3L visible at 4.3, 8.6, and 12.97 kHz, respectively. The increase in 1L mode frequency in the 200K case is attributed to a higher speed of sound in the unstable case.

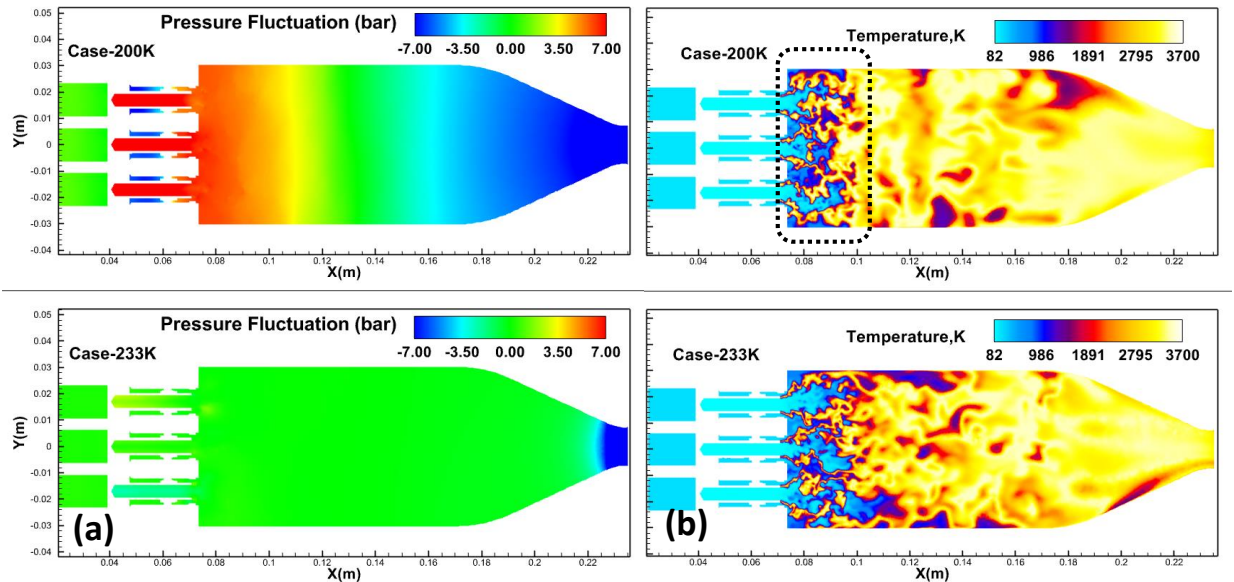

**FIG.34.** Comparison of (a) instantaneous pressure (b) temperature variation at the cut plane. Watch the animation for more details (*Multimedia view*)

## G. Instability mechanism

The raw pressure oscillations signify the onset of instability in the 200K case and are confirmed by detailed spectral analysis. The frequency analysis showcases the high amplitude pressure oscillations in the combustor's first longitudinal mode. The following section explores the instability mechanism, highlighting the possibility of combustion delay due to lower-temperature fuel injection. The higher potential of the feedback loop between unsteady pressure and heat release is discussed through instantaneous variation of major flow and flame parameters at an axial cut plane. Figure 34(a) reports the instantaneous pressure contour at a simulation time of 80ms (*Multimedia view*). It shows a spatial variation of pressure in 200K (top) and 233K (bottom) simulations. The 200K case clearly shows the formation of a high-pressure zone close to the injector faceplate and lower pressure at the opposite end. The observed pressure variation in 200K is characteristic of a longitudinal acoustic wave, while the 233K case shows negligible pressure variation at the axial cut plane. Figure 34(b) compares temperature variation for both cases. It shows a compressed flame zone (highlighted) near the injector outlet compared to the stable 233K case. It reveals the effect of longitudinal wave movement on the flame front, which gets distorted by incoming pressure waves. Cycle analysis helps to understand the instantaneous variation of parameters at different time intervals over one cycle time. A view of pressure fluctuation in the combustor at different time instants over an instability cycle is presented in Figure 35 (a & b). The pressure fluctuation plot at P1 and Pend with time instants (t1 to t5) marked is displayed in Figure 35(a), whereas the acoustic wave movement for the 200K case is depicted in Figure 35(b). The temporal pressure variation displays characteristics of a typical longitudinal acoustic wave in the cavity with high- and low-pressure regions at a fixed location over time.

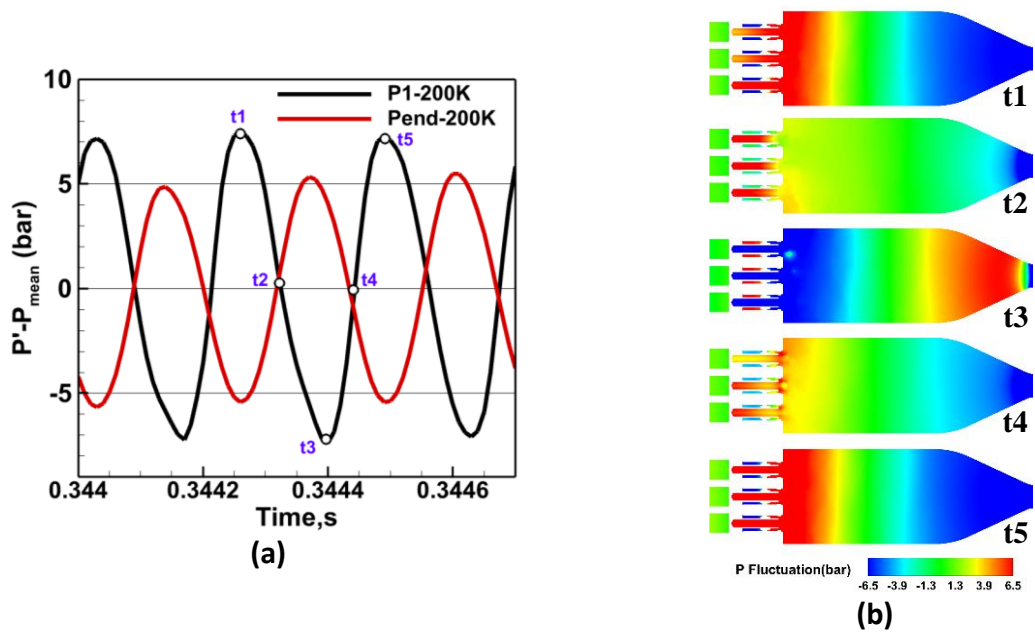

**FIG.35.** Cycle analysis: (a) pressure probe, (b) instantaneous pressure contour

The self-sustained nature of the acoustic wave seen in the 200K cases can be a collective effect of fuel-rich conditions in the back step region, a non-linear rise of methane-specific heat close to critical point temperature (190K) leading to higher thermal gradients in the shear layer and resultant hydrodynamic disturbances. Temporal vorticity statistics are collected at an injector exit location to understand the coupling between vorticity magnitude and pressure dynamics observed in the combustor for the 200K case. Figure 36(a) shows a variation of vorticity magnitude with intermittent spikes in the time domain. The FFT plot is displayed in Figure 36(b), which shows a dominant frequency of 4.3 kHz. The vorticity frequency coincides with the first longitudinal mode of the combustor, which indicates a frequency coupling. A positive feedback loop between vorticity generation and acoustic modes can explain the amplification of pressure fluctuations in the 200K case.

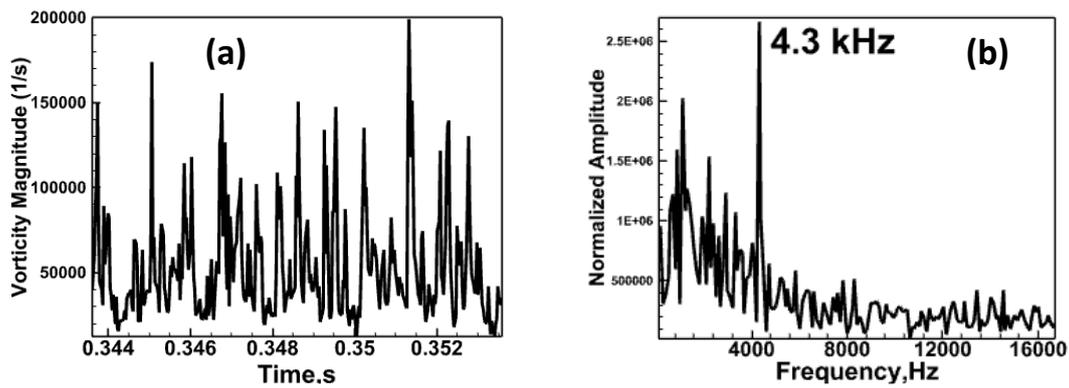

**FIG.36.** (a) Vorticity magnitude plot, (b) FFT of the vorticity magnitude spectrum

It is well known that pressure fluctuations are the manifestation of fluctuating heat release. The heat release fluctuations can be directly correlated to fluctuating fuel flow rates. The coupling between pressure, heat release, and fuel flow rate fluctuations is a fundamental characteristic of combustion dynamics. Figure 37(a) presents rapid fluctuations in methane mass fraction observed near the injector outlet (location marked).

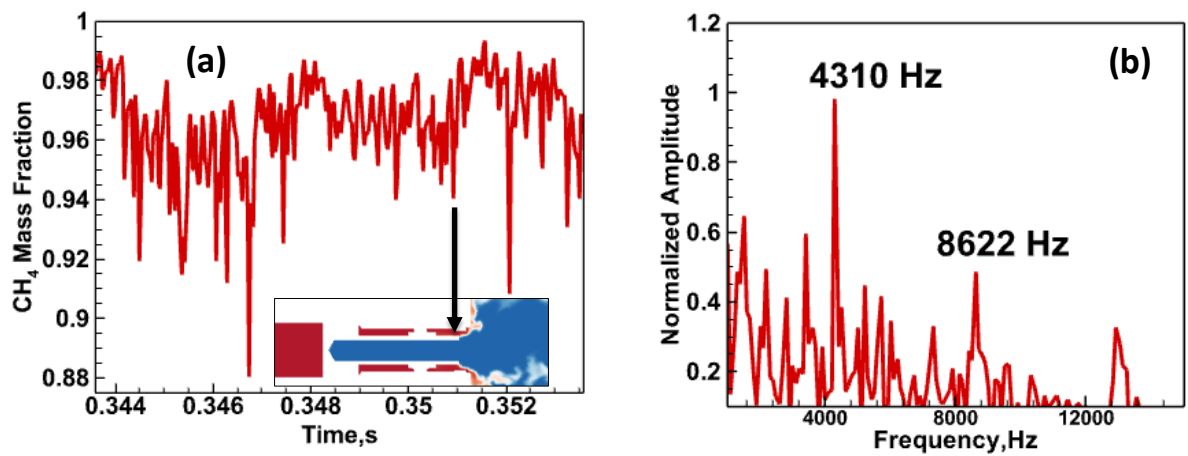

**FIG.37.** (a) Methane mass fraction plot, (b) FFT of mass fraction spectrum

The instantaneous methane mass fraction varies violently due to incoming pressure waves between 1 and 0.88. Figure 37(b) depicts a spectral plot of methane mass fraction, which shows two dominant frequency peaks, the same as the longitudinal oscillation modes of the combustor. The longitudinal mode pressure oscillations in the 200K case can result in fuel mass flow rate fluctuations, affecting the local mixture ratio (O/F), flame position, flame intensity, and flame propagation. The fluctuations in mass flow rate can perturb the heat release rate, which eventually influences and intensifies pressure oscillations.

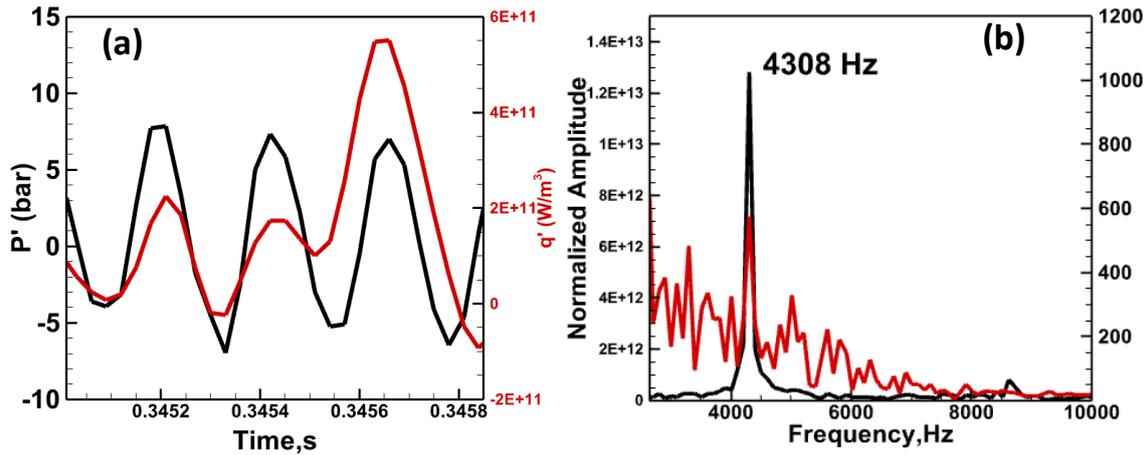

**FIG.38.** (a) $P'q'$ temporal variation, (b) FFT of $P'q'$ spectrum for 200K case

The in-phase pressure and heat release variation create a feedback loop sustaining instability. Figure 38(a) depicts the synchronized (in-phase) variation of instantaneous pressure and heat release over time. Figure 38(b) reveals a frequency match of instantaneous pressure and heat release rate, showing the presence of a feedback mechanism in the 200K case.

Instantaneous contours and probe data analysis reveal that longitudinal waves can induce instabilities in flames, especially when the acoustic wave frequency aligns with a resonant frequency of the flow and flame parameters. This resonance can result in self-sustained oscillations, as observed in the case of methane injection at 200K. Overall, this work elaborates on injection and combustion dynamics in our designed seven-element combustor and reveals multiple aspects related to the stability of LOx-methane supercritical combustion.

## V. CONCLUSION

A comprehensive attempt is made to decipher the complex combustion dynamic mechanisms and to gain improved insights into the stability aspects of multi-element LOx-methane rocket scale combustors. The work utilizes a high-fidelity numerical framework with real gas thermodynamics, high-pressure transport physics, and flamelet generated manifold for combustion closure. The methodology is validated on a benchmark Mascotte G2 test case with LOx-methane propellants, accurately reproducing transcritical oxygen injection and the supercritical flame structure. The validation study reveals the FGM model to accurately represent the high-pressure LOx-methane diffusion flame. Later, the validated methodology is applied to 7 bi-directional swirl coaxial injector combustor. The predictions adequately capture multi-element rocket combustor flow, flame interactions, and major features through detailed analysis. The predictions capture the movement of pressure waves in the combustor with an evident tangential acoustic mode. Detailed spectral analysis further identifies the dominant hydrodynamic and acoustic modes present in the combustor while revealing the effect of local sound speed on the evolution of transverse acoustic modes. The study further explores the interplay between injector and combustor dynamics, highlighting the dominance of longitudinal mode pressure oscillations in the injector LOx path. In contrast, near-wall injectors are influenced by tangential acoustic waves set up in the combustor, which can periodically affect pressure oscillations and lead to mass flow rate fluctuations in the injectors. The injector and chamber exhibit a dominant acoustic frequency of around 4 kHz, potentially leading to a frequency-coupled system and enhanced combustor dynamics. Our study on the effect of fuel injection temperature on combustor stability reveals violent dynamic activity at lower fuel injection temperatures. It captures the combustor's unstable longitudinal acoustic modes (1L and harmonics) and reproduces self-sustained limit cycle oscillations. The study showcases the onset of instability at lower methane injection temperatures, which can serve as a key observation for the design of a stable LOx-methane rocket engine. Overall, this study sheds light on the flow and flame dynamics in a multi-element rocket scale combustor. Further, it demonstrates the capability of numerical framework to capture fundamental physics associated with swirl injection, LOx-methane combustion dynamics, and the effect of lower fuel injection temperature on the stability of the combustor. The developed framework will be further tested to assess the combustor stability at perturbed/off-nominal flow rate conditions and varied injector recess to determine the feasibility of stable operation at lower fuel injection temperatures.

## ACKNOWLEDGEMENTS

The authors acknowledge the availability of 100 TeraFlop High-Performance Computing Facility at the Liquid Propulsion Systems Center (LPSC), featuring direct water-cooled processor technology. We thank M/s StarOneIT, Trivandrum, for their logistics support in maintaining this facility. Additionally, we acknowledge the technical assistance provided by the ANSYS India team.

# AUTHOR DECLARATIONS

**CONFLICT OF INTEREST**

The authors have no conflicts to disclose.

**DATA AVAILABILITY**

The data that support the findings of this study are available from the corresponding author upon reasonable request.

**APPENDIX A:** Favre filtered governing equations, departure function formalism, and FGM model details

**Filtered continuity:**

$$\frac{\partial \bar{\rho}}{\partial t} + \frac{\partial}{\partial x_i}(\bar{\rho}\tilde{u}_i) = 0 \tag{5}$$

**Filtered momentum:**

$$\frac{\partial}{\partial t}(\bar{\rho}\tilde{u}_i) + \frac{\partial}{\partial x_j}(\bar{\rho}\tilde{u}_i\tilde{u}_j) = \frac{\partial}{\partial x_j}(\widetilde{\sigma_{ij}}) - \frac{\partial \bar{p}}{\partial x_i} - \frac{\partial \tau_{ij}}{\partial x_j} \tag{6}$$

Where $\sigma_{ij}$ is the stress tensor due to molecular viscosity defined by:

$$\sigma_{ij} \equiv \left[\mu\left(\frac{\partial \bar{u}_i}{\partial x_j} + \frac{\partial \bar{u}_j}{\partial x_i}\right)\right] - \frac{2}{3}\mu \frac{\partial \bar{u}_l}{\partial x_l}\delta_{ij} \tag{7}$$

The compressible form of the subgrid stress tensor is defined as:

$$\tau_{ij} \equiv \bar{\rho}\widetilde{u_i u_j} - \bar{\rho}\tilde{u}_i\tilde{u}_j \tag{8}$$

**Energy/Enthalpy equation:**

$$\frac{\partial \bar{\rho}\tilde{h}_s}{\partial t} + \frac{\partial \bar{\rho}\tilde{u}_i\tilde{h}_s}{\partial x_i} - \frac{\partial \bar{p}}{\partial t} - \tilde{u}_j\frac{\partial \bar{p}}{\partial x_i} - \frac{\partial}{\partial x_i}\left(\lambda\frac{\partial \tilde{T}}{\partial x_i}\right) = -\frac{\partial}{\partial x_j}\left[\bar{\rho}(\widetilde{u_i h_s} - \tilde{u}_i\tilde{h}_s)\right] \tag{9}$$

Where $h_s$ & $\lambda$ are the sensible enthalpy and thermal conductivity, respectively. The subgrid enthalpy flux term in the filtered energy equation is approximated using the gradient hypothesis:

$$\bar{\rho}(\widetilde{u_i h_s} - \tilde{u}_i\tilde{h}_s) = -\frac{\mu_{SGS} c_p}{Pr_{SGS}}\frac{\partial \tilde{T}}{\partial x_j} \tag{10}$$

$\mu_{SGS}$ is the subgrid viscosity and $Pr_{SGS}$ is the subgrid Prandtl number and is equal to 0.85. The subgrid scale stress resulting from the filtering operation is modeled using sub-grid scale models. The Smagorinsky-Lilly model[47] based on the Bousinessq hypothesis is used to model sub-grid scale stress term, $\tau_{ij}$, where $\bar{S}_{ij}$ is the rate-of-strain tensor for the resolved scale.

$$\tau_{ij} = -2\mu_t \bar{S}_{ij} \tag{11}$$

$$\bar{S}_{ij} \equiv \frac{1}{2}\left(\frac{\partial \bar{u}_i}{\partial x_j} + \frac{\partial \bar{u}_j}{\partial x_i}\right) \tag{12}$$

A modified configuration for compressible flows is used:

$$\tau_{ij} = \tau_{ij} - \frac{1}{3}\tau_{kk}\delta_{ij} + \frac{1}{3}\tau_{kk}\delta_{ij} \tag{13}$$

$(\tau_{ij} - \frac{1}{3}\tau_{kk}\delta_{ij})$ is a deviatoric part of subgrid-scale stress and $\frac{1}{3}\tau_{kk}\delta_{ij}$ is an isotropic part. The isotropic part of the subgrid-scale stresses is not modeled but added to the filtered static pressure term [47]. The deviatoric part of the subgrid-scale stress tensor is modeled using a compressible form of the Smagorinsky model:

$$\tau_{ij} - \frac{1}{3}\tau_{kk}\delta_{ij} = -2\mu_t \left(S_{ij} - \frac{1}{3}S_{kk}\delta_{ij}\right) \tag{14}$$

The eddy viscosity is modeled by the Smagorinsky-Lilly model given as:

$$\mu_t = \rho L_s^2 |\bar{S}| \tag{15}$$

Where $L_s$ is the mixing length for the subgrid scales and

$$|\bar{S}| = \sqrt{2\bar{S}_{ij}\bar{S}_{ij}} \tag{16}$$

$L_s$ is computed using, $L_s = \min(kd, C_s\Delta)$ where $C_s$ is Smagorinsky constant and $\Delta$ is the local grid scale. The dynamic version of the Smagorinsky-Lilly model by Germano et al.[47] is used in this study in which $C_s$ is dynamically computed based on the information provided by the resolved scales of motion.

**Departure function formalism:**

The internal energy $e$, enthalpy $h$, entropy $S$, and specific heat $C_P$ are calculated by this formalism and are described as follows:

$$e(T,\rho) = e_0(T) + \int_{\rho_0}^{\rho} \left[\frac{P}{\rho^2} - \frac{T}{\rho^2}\left(\frac{\partial P}{\partial T}\right)_\rho\right]_T d\rho \tag{17}$$

$$h(T,P) = h_0(T) + \int_{P_0}^{P} \left[\frac{1}{\rho} - \frac{T}{\rho^2}\left(\frac{\partial \rho}{\partial T}\right)_P\right]_T dP \tag{18}$$

$$S(T,P) = S_0(T) - \int_{\rho_0}^{\rho} \left[\frac{1}{\rho^2}\left(\frac{\partial P}{\partial T}\right)_\rho\right]_T d\rho \tag{19}$$

$$C_P(T,\rho) = C_{P_0}(T) - \int_{\rho_0}^{\rho} \left[\frac{T}{\rho^2}\left(\frac{\partial^2 P}{\partial T^2}\right)_\rho\right]_T d\rho + \frac{T}{\rho^2} \frac{\left(\frac{\partial^2 P}{\partial T^2}\right)_\rho}{\left(\frac{\partial P}{\partial \rho}\right)_T} \tag{20}$$

Where $\rho$ is the fluid density at pressure, $P$ and temperature, $T$ condition. The subscript 0 represents the ideal state, and the departure functions on the right-hand side are determined using a real fluid equation of state.

**FGM formulation and transport equations:**

The FGM formulation uses the reaction progress variable $Y_c$ as manifold control variable, which is expressed in terms of species mass fraction, $Y_k$, where k is the species index, and $a_k$ is a weight constant. The unburnt mass fraction of species is depicted with superscript u, which accounts for any product species present in the oxidizer or fuel stream due to lean premixed combustion, pilot fuel inlets, or an inert stream in the oxidizer. The reaction progress variable is given as follows:

$$Y_c = \sum_k a_k (Y_k - Y_k^u) \tag{21}$$

The progress variable definition follows monotonic variation from unburnt to burnt state, with equal species weights for $CO_2$ and $CO$, used as 1 in this study. Equal weights are used in progress variable definition to capture different time scales embedded in the complete reaction manifold. Yadav et al.[53] described the appropriateness of equal species weight for defining progress variable for lifted flame simulations. This formulation uses a normalized version of the progress variable to optimize the performance of pre-processed look-up tables. The normalized progress variable varies from 0 in the unburnt to 1 in the burnt zone. The progress variable $c$ is the normalized sum of the product species mass fractions. It is expressed as the ratio of the reaction progress variable of product species at any instant to its value at equilibrium given by:

$$c = \frac{\sum_k a_k (Y_k - Y_k^u)}{\sum_k a_k (Y_k^{eq} - Y_k^u)} = \frac{Y_c}{Y_c^{eq}} \tag{22}$$

where, $Y_c^{eq}$ is the equilibrium value and is a function of the local mixture fraction only. A parametric study by Yadav et al.[53,54] shows that normalization with low strain rate flamelet composition does not significantly differ from normalization with equilibrium progress variable.

In the FGM framework, the transport equations for both control variables, i.e., mixture fraction $f$, and the un-normalized progress variable, $Y_c$ is solved along with basic governing equations. The Favre averaged transport equation for the mixture fraction, $\tilde{f}$ is given by:

$$\frac{\partial}{\partial t}(\bar{\rho}\tilde{f}) + \nabla \cdot (\bar{\rho}\vec{v}\tilde{f}) = \nabla \cdot \left(\frac{\mu_{eff}}{\sigma_t} \nabla \tilde{f}\right) + S_m \tag{23}$$

where, $S_m$ is only due to the mass transfer into the gas phase from liquid droplets. $\mu_{eff}$ is the effective viscosity composed of a laminar ($\mu_l$) and a turbulent contribution ($\mu_t$). In this diffusion flame framework, the transport equation for density weighted un-normalized progress variable, $\tilde{Y}_c$ is solved, instead of normalized progress variable, $c$ as $c$ transport equation has additional terms and is not well defined at the oxidizer boundaries. Moreover, the solution of $Y_c$ brings added advantage to model the flame quenching naturally with its definition.

$$\frac{\partial \bar{\rho}\tilde{Y}_c}{\partial t} + \frac{\partial}{\partial x_i}(\bar{\rho}\tilde{u}_i \tilde{Y}_c) = \frac{\partial}{\partial x_i}\left(D_{eff} \frac{\partial \tilde{Y}_c}{\partial x_i}\right) + \bar{S}_{Y_c} \tag{24}$$

In this FGM formulation, $\bar{S}_{Y_c}$ is modeled as a finite rate source term taken from the flamelet library using $c$ and $f$ as control variables. This means source term $\bar{S}_{Y_c}$ determines the turbulent flame position and is calculated as:

$$\bar{S}_{Y_c} = \bar{\rho} \iint S_{FR}(c,f) p(c,f) \, dc \, df = \bar{S}_{FR} \tag{25}$$

The $p(c,f)$ is a joint probability density function (PDF) specified as the product of two beta PDFs. Our previous work[1] provides a mathematical description of beta PDF, which requires information on the variance of scalar fields to calculate mean quantities. In this study, the variance of the un-normalized reaction progress variable is modeled using the algebraic formulation, assuming the equilibrium of the generation and dissipation of variance at the sub-grid scale.

$$\widetilde{Y_c''^2} = C_{var} \frac{l_{turb}^2}{Sc_t} (\nabla \widetilde{Y_c})^2 \tag{26}$$

Where $l_{turb}$ is the turbulent length scale and $C_{var}$ is the constant with a value assigned as 0.1 in this formulation. The turbulent Schmidt number is used as 0.7 in this study. Similarly, the transport equation for mixture fraction variance is not solved; instead, it is modeled by an algebraic equation given as:

$$\overline{f'^2} = C_{var} L_s^2 |\nabla \bar{f}|^2 \tag{27}$$

The constant $C_{var}$ is computed dynamically based on the dynamic version of the Smagorinsky-Lilly SGS model used in this study. Where $L_s$ is the mixing length for the subgrid scales and is computed using, $L_s = \min(kd, C_s \Delta)$ where $C_s$ is Smagorinsky constant and $\Delta$ is the local grid scale. The constant $C_s$ is also computed dynamically based on the information provided by the resolved scales of motion.

This section highlights the turbulence-chemistry interaction in the FGM model: The shape of the PDF is computed from the mean and variance of the mixture fraction, progress variable, and enthalpy at each point in the flow field. The calculation of mean scalars involves the integration of the instantaneous (flamelet profiles) values with a Favre presumed probability density function (PDF). This formulation generates adiabatic flamelets, assuming a negligible heat gain/loss effect on the species mass fractions. Although a non-adiabatic version of the PDF table is generated with an extra dimension of mean enthalpy $\bar{H}$ to consider the effect of enthalpy loss or gain on the scalar computation. A non-adiabatic energy treatment is employed with a multi-dimensional look-up table generation at 21 enthalpy levels, above and below the adiabatic enthalpy level. The density-weighted mean scalars (such as species fractions and temperature), denoted by $\bar{\phi}$, are calculated as:

$$\bar{\phi} = \iint_0^1 \phi(f, c, \bar{H}) p(f, c)\, df\, dc \tag{28}$$

where $\phi = \rho, T, C_p, Y_k, \bar{S}_{FR}$ where $\rho$ is mean density, $T$ is temperature, $C_p$ is specific heat, $Y_k$ is species mass fraction, and $\bar{S}_{FR}$ is mean finite rate source term.

# APPENDIX B: Pressure and Frequency Comparison in 233K and 200K Simulations:

This section compares pressure and frequency in 233K and 200K simulations. Figure 39 compares the frequency spectrum for all P1 probes, i.e. P1, P1c, and P1d, for 200K and 233K cases. All P1 probes display a high amplitude frequency peak compared to the 233K case, although both cases display the same amplitude for frequency corresponding to the 1T mode of the combustor at 8 kHz. The P1c probe does not show the third harmonic frequency peak, as seen in the P1 and P1d frequency spectra.

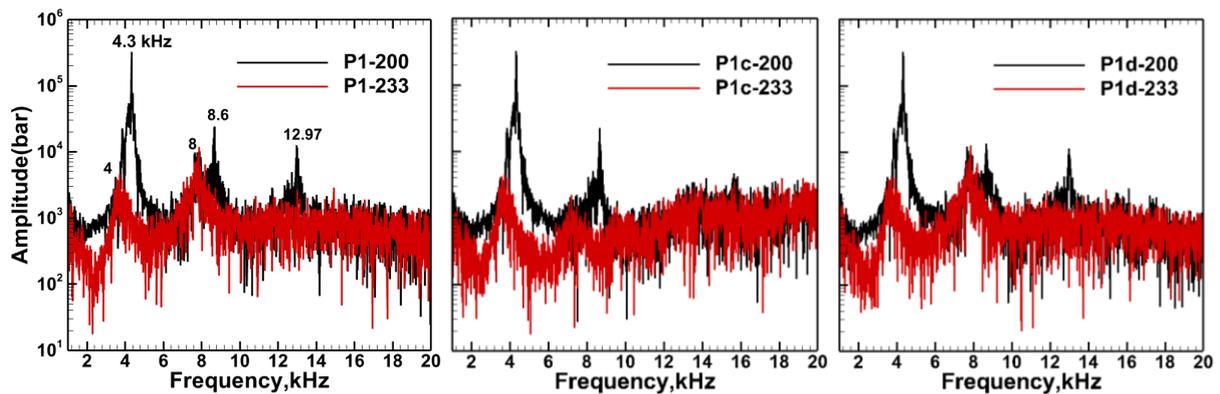

FIG.39. FFT comparison of all P1 probes in 200K and 233K case

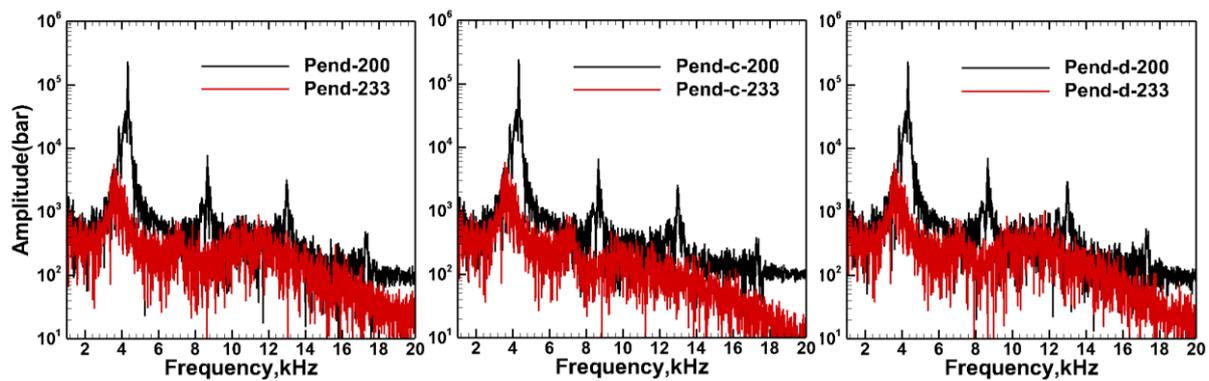

FIG.40 FFT comparison of all Pend probes in 200K and 233K case

A similar FFT variation is seen for Pend probes in Figure 40, which also displays high amplitude peaks in the unstable 200K case and stable characteristics in the 233K case. It is noticed that the central probe at the end location labeled as Pend-c displays a third harmonic of 1L. The end probes also display the fourth harmonic peak at a frequency close to 17 kHz. The amplitude of 1L mode in the unstable case is identical at P1 and Pend probe locations.

We plot the frequency spectrum separately at each probe station to identify the transverse mode in an unstable 200K case (1, 2, 3, end). Figure 41 displays the FFT spectrum of all probe stations for the 200K case. The frequency plot reveals the presence of transverse mode (1T) at 8 kHz in all near-wall probes (marked with an arrow) except at the end location (Pend) probe plot. The center probes P1c, P2c, and P3c also do not exhibit any transverse wave movement. It is well known that the transverse mode develops close to the injector faceplate and combustor walls and is correctly reproduced in this study.

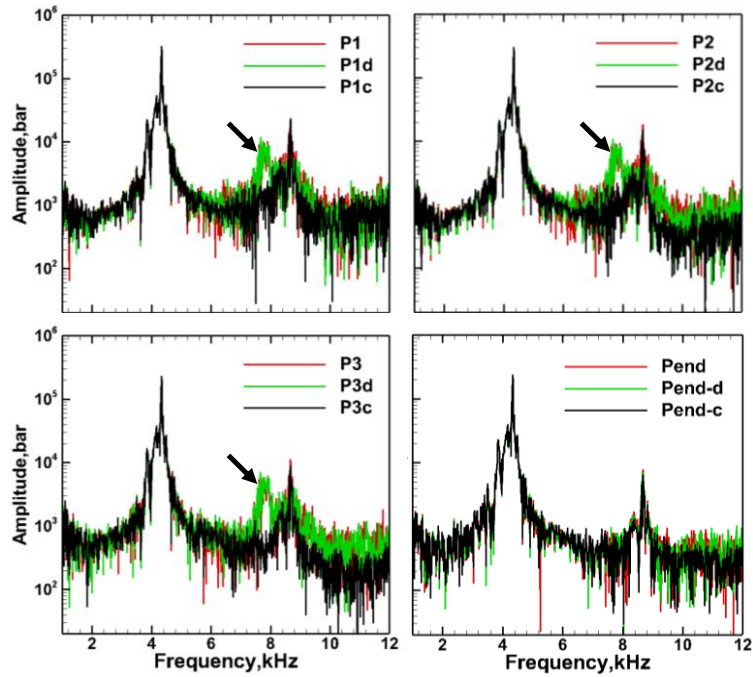

**FIG.41.** FFT spectrum of all probes in 200K case

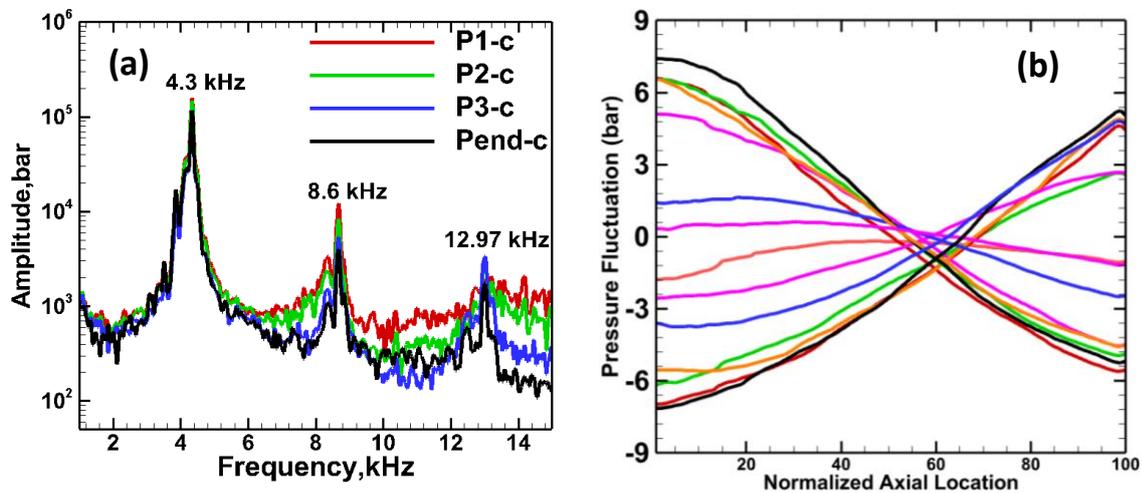

**FIG.42.** (a) FFT spectrum of all center probes in 200K case, (b) Wave movement in the combustor

As the central probes are not influenced by transverse wave propagation, all center probe data can clearly show the triggered longitudinal modes in the 200K case. Figure 42(a, b) displays the triggered modes and acoustic wave movement inside the combustor for the 200K case. Figure 42(a) clearly shows the triggered modes as 1L and harmonics, with maximum acoustic energy possessed by 1L mode. The lower frequency mode (1L: 4.3kHz) dominates as the higher frequency modes tend to experience more significant losses due to viscous damping. The longitudinal wave movement in the combustor is depicted in Figure 42 (b), which shows pressure variation along the axial length of the combustor at different time intervals. It shows the pressure variation at a line along the axis of the combustor at different time instants. The pressure variation at different time instants represents high-pressure longitudinal wave movement in the 200K case.